\documentclass[10pt,journal,letterpaper,compsoc]{IEEEtran}
\ifCLASSOPTIONcompsoc
  \usepackage[nocompress]{cite}
\else
  \usepackage{cite}
\fi
\ifCLASSINFOpdf
  \usepackage[pdftex]{graphicx}
  \DeclareGraphicsExtensions{.pdf,.jpeg,.png}
\else
  \usepackage[dvips]{graphicx}
  \DeclareGraphicsExtensions{.eps}
\fi
\usepackage[cmex10]{amsmath}
\usepackage{amssymb}
\usepackage{bm}
\newcommand{\transpose}{^{\mathrm T} }

\usepackage{threeparttable}

\ifCLASSOPTIONcompsoc
  \usepackage[caption=false,font=normalsize,labelfont=sf,textfont=sf]{subfig}
\else
  \usepackage[caption=false,font=footnotesize]{subfig}
\fi
\usepackage{stfloats}
\usepackage{url}
\begin{document}
%
% paper title
% can use linebreaks \\ within to get better formatting as desired
\title{Negative Example Aided Transcription Factor Binding Site Search}
%
%
% author names and IEEE memberships
% note positions of commas and nonbreaking spaces ( ~ ) LaTeX will not break
% a structure at a ~ so this keeps an author's name from being broken across
% two lines.
% use \thanks{} to gain access to the first footnote area
% a separate \thanks must be used for each paragraph as LaTeX2e's \thanks
% was not built to handle multiple paragraphs
%
%
%\IEEEcompsocitemizethanks is a special \thanks that produces the bulleted
% lists the Computer Society journals use for "first footnote" author
% affiliations. Use \IEEEcompsocthanksitem which works much like \item
% for each affiliation group. When not in compsoc mode,
% \IEEEcompsocitemizethanks becomes like \thanks and
% \IEEEcompsocthanksitem becomes a line break with idention. This
% facilitates dual compilation, although admittedly the differences in the
% desired content of \author between the different types of papers makes a
% one-size-fits-all approach a daunting prospect. For instance, compsoc 
% journal papers have the author affiliations above the "Manuscript
% received ..."  text while in non-compsoc journals this is reversed. Sigh.

\author{Chih~Lee %,~\IEEEmembership{Member,~IEEE,}
        %John~Doe,~\IEEEmembership{Fellow,~OSA,}
        and~Chun-Hsi~Huang%,~\IEEEmembership{Life~Fellow,~IEEE}% <-this % stops a space
\IEEEcompsocitemizethanks{
%\IEEEcompsocthanksitem M. Shell is with the Department
%of Electrical and Computer Engineering, Georgia Institute of Technology, Atlanta,
%GA, 30332.\protect\\
% note need leading \protect in front of \\ to get a newline within \thanks as
% \\ is fragile and will error, could use \hfil\break instead.
%E-mail: see http://www.michaelshell.org/contact.html
\IEEEcompsocthanksitem C. Lee and C.-H. Huang are with the Department
of Computer Science and Engineering, University of Connecticut, Storrs,
CT, 06269.\protect\\
E-mail: \{chihlee,huang\}@engr.uconn.edu}% <-this % stops a space
\thanks{}}

% note the % following the last \IEEEmembership and also \thanks - 
% these prevent an unwanted space from occurring between the last author name
% and the end of the author line. i.e., if you had this:
% 
% \author{....lastname \thanks{...} \thanks{...} }
%                     ^------------^------------^----Do not want these spaces!
%
% a space would be appended to the last name and could cause every name on that
% line to be shifted left slightly. This is one of those "LaTeX things". For
% instance, "\textbf{A} \textbf{B}" will typeset as "A B" not "AB". To get
% "AB" then you have to do: "\textbf{A}\textbf{B}"
% \thanks is no different in this regard, so shield the last } of each \thanks
% that ends a line with a % and do not let a space in before the next \thanks.
% Spaces after \IEEEmembership other than the last one are OK (and needed) as
% you are supposed to have spaces between the names. For what it is worth,
% this is a minor point as most people would not even notice if the said evil
% space somehow managed to creep in.

% The paper headers
\markboth{Journal of \LaTeX\ Class Files,~Vol.~6, No.~1, January~2007}%
{Lee and Huang: Negative Example Aided TFBS Search}
% The only time the second header will appear is for the odd numbered pages
% after the title page when using the twoside option.
% 
% *** Note that you probably will NOT want to include the author's ***
% *** name in the headers of peer review papers.                   ***
% You can use \ifCLASSOPTIONpeerreview for conditional compilation here if
% you desire.

% The publisher's ID mark at the bottom of the page is less important with
% Computer Society journal papers as those publications place the marks
% outside of the main text columns and, therefore, unlike regular IEEE
% journals, the available text space is not reduced by their presence.
% If you want to put a publisher's ID mark on the page you can do it like
% this:
%\IEEEpubid{0000--0000/00\$00.00~\copyright~2007 IEEE}
% or like this to get the Computer Society new two part style.
%\IEEEpubid{\makebox[\columnwidth]{\hfill 0000--0000/00/\$00.00~\copyright~2007 IEEE}%
%\hspace{\columnsep}\makebox[\columnwidth]{Published by the IEEE Computer Society\hfill}}
% Remember, if you use this you must call \IEEEpubidadjcol in the second
% column for its text to clear the IEEEpubid mark (Computer Society jorunal
% papers don't need this extra clearance.)

% for Computer Society papers, we must declare the abstract and index terms
% PRIOR to the title within the \IEEEcompsoctitleabstractindextext IEEEtran
% command as these need to go into the title area created by \maketitle.
\IEEEcompsoctitleabstractindextext{%
\begin{abstract}
%\boldmath
Computational approaches to transcription factor binding site identification have been actively
researched for the past decade.
Negative examples have long been utilized in \textit{de novo} motif discovery
and have been shown useful in transcription factor binding site search as well.
However, understanding of the roles of negative examples in binding site search is still very limited.

We propose the 2-centroid and optimal discriminating vector methods, taking into account negative examples.
Cross-validation results on \textit{E. coli} transcription factors show that the proposed methods benefit
from negative examples, outperforming the centroid and position-specific scoring matrix methods.
We further show that our proposed methods perform better than a state-of-the-art method.
We characterize the proposed methods in the context of the other compared methods and show that,
coupled with motif subtype identification, the proposed methods can be effectively applied to
a wide range of transcription factors. Finally, we argue that the proposed methods are well-suited
for eukaryotic transcription factors as well.

Software tools are available at: \url{http://biogrid.engr.uconn.edu/tfbs_search/}.

\end{abstract}
% IEEEtran.cls defaults to using nonbold math in the Abstract.
% This preserves the distinction between vectors and scalars. However,
% if the journal you are submitting to favors bold math in the abstract,
% then you can use LaTeX's standard command \boldmath at the very start
% of the abstract to achieve this. Many IEEE journals frown on math
% in the abstract anyway. In particular, the Computer Society does
% not want either math or citations to appear in the abstract.

% Note that keywords are not normally used for peer review papers.
}

% make the title area
\maketitle

% To allow for easy dual compilation without having to reenter the
% abstract/keywords data, the \IEEEcompsoctitleabstractindextext text will
% not be used in maketitle, but will appear (i.e., to be "transported")
% here as \IEEEdisplaynotcompsoctitleabstractindextext when compsoc mode
% is not selected <OR> if conference mode is selected - because compsoc
% conference papers position the abstract like regular (non-compsoc)
% papers do!
\IEEEdisplaynotcompsoctitleabstractindextext
% \IEEEdisplaynotcompsoctitleabstractindextext has no effect when using
% compsoc under a non-conference mode.

% For peer review papers, you can put extra information on the cover
% page as needed:
% \ifCLASSOPTIONpeerreview
% \begin{center} \bfseries EDICS Category: 3-BBND \end{center}
% \fi
%
% For peerreview papers, this IEEEtran command inserts a page break and
% creates the second title. It will be ignored for other modes.
\IEEEpeerreviewmaketitle

\section{Introduction}
% Computer Society journal papers do something a tad strange with the very
% first section heading (almost always called "Introduction"). They place it
% ABOVE the main text! IEEEtran.cls currently does not do this for you.
% However, You can achieve this effect by making LaTeX jump through some
% hoops via something like:
%
%\ifCLASSOPTIONcompsoc
%  \noindent\raisebox{2\baselineskip}[0pt][0pt]%
%  {\parbox{\columnwidth}{\section{Introduction}\label{sec:introduction}%
%  \global\everypar=\everypar}}%
%  \vspace{-1\baselineskip}\vspace{-\parskip}\par
%\else
%  \section{Introduction}\label{sec:introduction}\par
%\fi
%
% Admittedly, this is a hack and may well be fragile, but seems to do the
% trick for me. Note the need to keep any \label that may be used right
% after \section in the above as the hack puts \section within a raised box.

% The very first letter is a 2 line initial drop letter followed
% by the rest of the first word in caps (small caps for compsoc).
% 
% form to use if the first word consists of a single letter:
% \IEEEPARstart{A}{demo} file is ....
% 
% form to use if you need the single drop letter followed by
% normal text (unknown if ever used by IEEE):
% \IEEEPARstart{A}{}demo file is ....
% 
% Some journals put the first two words in caps:
% \IEEEPARstart{T}{his demo} file is ....
% 
% Here we have the typical use of a "T" for an initial drop letter
% and "HIS" in caps to complete the first word.
%TRANSCRIPTIONAL REGULATION
\IEEEPARstart{T}{ranscription} of genes followed by translation of their transcripts into proteins determines the type
and functions of a cell. Expression of certain genes even initiates or suppresses differentiation of
stem cells. It is therefore crucial to understand the mechanisms of transcriptional regulation.
%TRANSCRIPTIONAL REGULATION BY TF BINDING
Among them, transcription factor (TF) binding is the one that has been given considerable attention by
computational biologists for the past decade and is still being actively researched. A TF is a
protein or protein complex that regulates transcription of one or more genes by binding to the double-stranded
DNA. A first step in computational identification of target genes regulated by a TF is to pinpoint its binding
sites in the genome. Once the binding sites are found, the putative target genes can be searched and located
in flanking regions of the binding sites.

%TFBS DISCOVERY VERSUS TFBS SEARCH PROBLEMS
In general, there are two approaches to computational transcription factor binding site (TFBS)
identification, motif discovery and TFBS search. The former assumes that a set of sequences is given
and each of the sequences may or may not contain TFBS's. An algorithm then predicts the locations and
lengths of TFBS's. The term motif refers to the pattern that are shared by the discovered TFBS's.
This kind of algorithms relies on no prior knowledge of the motif and hence is known as \textit{de novo}
motif discovery algorithms.
The latter assumes that, in addition to a set of sequences, the locations and
lengths of TFBS's are known. An algorithm then learns from these examples and predicts TFBS's in
new sequences. Such algorithms are also called supervised learning algorithms since they are guided by
the given sequences with known TFBS's.

%REVIEW OF PREVIOUS WORK
Plenty of efforts have been devoted to the \textit{de novo} motif discovery problem
\cite{Vil00,Bar01,Buh02,Sin02,Tak04,Raj05,Bal06,LiN06, Zas06,Yan09,Geo10}. Comprehensive evaluation and
comparison of the developed tools have been performed by Tompa \textit{et al.} \cite{Tom05} and Hu \textit{et al.} \cite{HuJ05}.
In this study, we focus on the problem of TFBS search. We refer readers interested in the motif
discovery problem to the evaluation and review articles \cite{Tom05,HuJ05,San06} and references therein.

A typical TFBS search method searches for the binding sites of a particular transcription factor in the following manner.
It scans a target DNA sequence and compare each $l$-mer to the binding site profile of the TF, where $l$ is the length
of a binding site. Each of the $l$-mer is scored when comparing to the profile. A cut-off score is then set by
the method to select candidate TF binding sites. The position-specific scoring matrix is a widely used profile
representation, where the binding sites of a TF are encoded as a $4 \times l$ matrix. Column $i$ of the matrix stores
the scores of matching the $i^{\text{th}}$ letter in an $l$-mer to nucleotides A, C, G and T, respectively.
Depending on the method of choice, the score of A at position $i$ can be the count of A at position $i$ in the
known TFBS's, the log-transformed probability of observing A at position $i$, or any other reasonable number.

Plenty of novel methods were based on this simple scoring method. 
Osada \textit{et al.} \cite{Osa04} extended this scoring approach by considering pairs of nucleotides
and weighting nuclueotide and nucleotide pairs by information content.
Extensive leave-one-out (LOO) cross-validation (CV) experiments were conducted on 35 TF's with
totally 410 binding sites. The results showed significant improvement regardless of the model used for
motif representation.
In a recent study, Salama and Stekel\cite{Sal10} showed correlations between two nucleotides within a TFBS by plotting
the mutual information matrix of a motif, reinforcing the findings reported in \cite{Osa04}.
A novel scoring method called the ungapped likelihood under positional background (ULPB) method was
proposed in this study. The ULPB method models a TFBS by two first-order Markov chains and scores
a candidate binding site by likelihood ratio produced by the two Markov chains.
LOO results on 22 TF's with 20 or more binding sites showed that ULPB is superior to the methods
compared in their work.

% WHY SHOULD NEGATIVE EXAMPLES BE USED
Explicit use of negative examples in the TFBS search problem is hindered by the vast amount of non-binding
sites of a transcription factor. This is further aggravated by the low specificity of some transcription factors,
where a binding site may be more similar to a non-binding site than some other binding sites.
Due to these issues, previous studies involving negative examples are limited and the roles of negative examples
remain unclear. In a review article, Hannenhalli \cite{Han08} surveyed work
on improved motif models and integrative methods. None of these reviewed studies \cite{Han08},
however, investigated the use of negative examples on top of true TFBS's.
While introducing improved benchmarks for computational motif discovery,
Sandve \textit{et al.} \cite{San07} described algorithms for finding optimal motif models using both positive and negative
TFBS's. Three models were compared using the proposed benchmarks. However, no methods relying on
only positive examples were compared.
Recently, Do and Wang \cite{DoH09} formulated the TFBS search problem as a classification problem,
proposed a novel similarity measure, and investigated three classification techniques.
Five-fold CV results showed that learning vector quantization performed better than
P-Match \cite{Che05}, which requires only positive examples. The evaluation, however, was done on
only 8 human transcription factors and 8 artificial ones.
It is not clear how the results on the small set of 8 real TF's can be related to other TF's.
%such as those curated by RegulonDB \cite{Gam08} and JASPAR \cite{Bry08}.

% needed in second column of first page if using \IEEEpubid
%\IEEEpubidadjcol

The goal of this study is to investigate the inclusion of negative examples in addition to
positive ones in TFBS search. We propose and characterize two novel extensions of the centroid method 
introduced in \cite{Osa04}. Besides the sequence similarity measures employed in
\cite{Osa04}, we also incorporate the novel similarity measure in \cite{DoH09} into an extension
of the centroid method. 
We compare our proposed methods to methods that do not rely upon negative examples, that is, the centroid
method, the ULPB method \cite{Sal10} and the well-known position-specific scoring matrix method.
Performance of a method is assessed by LOO CV experiments on two data sets of 35 and 26 transcription factors,
respectively. Moreoever, we discuss the situations when
the proposed methods can accurately differentiate binding sites from non-binding sites.
Advantages of coupling motif subtype identification with the proposed methods are also discussed.

The paper is organized as follows. In Section~\ref{sec:method}, we introduce existing methods compared in this
study and describe two novel methods proposed in this work. Leave-one-out cross-validation results on
two data sets are presented in Section~\ref{sec:results}. In Section~\ref{sec:discussion}, properties of
the proposed methods are studied and discussed. Connections between the proposed methods and the other compared
methods are established. Finally, we give the concluding remarks in Section~\ref{sec:conclusion}.

\section{Methods}\label{sec:method}

\subsection{Data sets}\label{sec:dataset}

For ease of comparison, we conduct experiments on two data sets used in previous work.% \citep{Osa04,Sal10}.
The first set was collected by Osada \textit{et al.} \cite{Osa04}, which
consists of 410 binding sites of 35 TF's with flanking regions located in the
\textit{E. coli} K-12 genome (version M54 of strain MG1655\cite{Bla97}). The statistics of
this data set are listed in Table~\ref{tab:dataset1}.
The second one also contains binding sites of TF's in the \textit{E. coli} K-12 genome
and was considered in \cite{Sal10}. We downloaded the latest data (release 6.8) from RegulonDB \cite{Gam08}
and kept only 26 TF's with 17 or more known binding sites. We summarize the data set in Table~\ref{tab:dataset2}.

\begin{table}[!t]
%% increase table row spacing, adjust to taste
%\renewcommand{\arraystretch}{1.3}
% if using array.sty, it might be a good idea to tweak the value of
% \extrarowheight as needed to properly center the text within the cells
\caption{Statistics of the first data set with 35 TF's}
\label{tab:dataset1}
\centering
%% Some packages, such as MDW tools, offer better commands for making tables
%% than the plain LaTeX2e tabular which is used here.
\begin{tabular}{lrrlrr}\hline
Name & Length & \# TFBS's & Name & Length & \# TFBS's\\\hline\hline
araC & 48 & 6 & arcA & 15 & 13\\
argR & 18 & 17 & cpxR & 15 & 12\\
crp & 22 & 49 & cspA & 20 & 4\\
cytR & 18 & 5 & dnaA & 15 & 8\\
fadR & 17 & 7 & fis & 35 & 19\\
fnr & 22 & 13 & fruR & 16 & 12\\
fur & 18 & 9 & galR & 16 & 7\\
gcvA & 20 & 4 & glpR & 20 & 13\\
hipB & 30 & 4 & ihf & 48 & 26\\
lexA & 20 & 19 & lrp & 25 & 14\\
malT & 10 & 10 & metJ & 16 & 15\\
metR & 15 & 8 & nagC & 23 & 6\\
narL & 16 & 10 & ntrC & 17 & 5\\
ompR & 20 & 9 & oxyR & 39 & 4\\
phoB & 22 & 15 & purR & 26 & 22\\
soxS & 35 & 14 & torR & 10 & 4\\
trpR & 24 & 4 & tus & 23 & 6\\
tyrR & 22 & 17 & & &\\\hline
\end{tabular}
\end{table}

\begin{table}[!t]
%% increase table row spacing, adjust to taste
%\renewcommand{\arraystretch}{1.3}
% if using array.sty, it might be a good idea to tweak the value of
% \extrarowheight as needed to properly center the text within the cells
\caption{Statistics of the second data set with 26 TF's}
\label{tab:dataset2}
\centering
%% Some packages, such as MDW tools, offer better commands for making tables
%% than the plain LaTeX2e tabular which is used here.
\begin{tabular}{lrrlrr}\hline
Name & Length & \# TFBS's & Name & Length & \# TFBS's\\\hline\hline
MetJ & 8 & 29 & Lrp & 12 & 62\\
SoxS & 18 & 19 & H-NS & 15 & 37\\
FlhDC & 16 & 20 & AraC & 18 & 20\\
Fis & 15 & 206 & ArcA & 15 & 93\\
IHF & 13 & 101 & OmpR & 20 & 22\\
PhoB & 20 & 17 & GlpR & 20 & 23\\
OxyR & 17 & 41 & CpxR & 15 & 37\\
NarL & 7 & 90 & CRP & 22 & 249\\
TyrR & 18 & 19 & NarP & 7 & 20\\
Fur & 19 & 81 & LexA & 20 & 40\\
NtrC & 17 & 17 & FNR & 14 & 87\\
MalT & 10 & 20 & PhoP & 17 & 21\\
ArgR & 18 & 32 & NsrR & 11 & 37\\\hline
\end{tabular}
\end{table}

\subsection{The centroid and 2-centroid methods}\label{sec:centroid}

We introduce the centroid method proposed by Osada \textit{et al.} \cite{Osa04} in a different manner.
We first define the similarity measure between two sequences $s$ and $t$ of length $l$.
\begin{align}
	\mathrm{Sim}(s, t) = \sum_{i = 1}^l w_i \mathcal{I}_{s_i}(t_i)\label{eq:sim},
\end{align}
where $s_i$ ($t_i$) is the $i^{\text{th}}$ letter of $s$ ($t$), $w_i$ denotes the weight on the
$i^{\text{th}}$ letter and $\mathcal{I}_{s_i}(\cdot)$ is the indicator function given by
\[
	\mathcal{I}_{s_i}(t_i) = \left\{ \begin{array}{cc}
		1 & \text{if }t_i = s_i,\\
		0 & \text{otherwise.}
	\end{array} \right.
\]
%, equals 1 if and only if $t_i = s_i$.
In this work, $w_i$ is set to either $1$ or the information content at position $i$ defined as
\begin{equation}\label{eq:IC}
	IC_i = 2 + \sum_{u \in \{\text{A, C, G, T}\}} f_i(u)\log_2\left[f_i(u)\right],
\end{equation}
where $f_i(u)$ is the probability of observing letter $u$ at position $i$.
When $w_i = 1$ for all $i$,
$\mathrm{Sim}(s, t)$ simply counts the number of letters shared between $s$ and $t$.
When pairs of nucleotides are taken into account, the similarity measure is defined as follows:
\begin{align}
	\mathrm{Sim2}(s, t) = \mathrm{Sim}(s, t)
	+ \sum_{k = 1}^K\sum_{i = 1}^{l-k} w_{i, j} \mathcal{I}_{s_is_{j}}(t_it_{j})\label{eq:sim2},
\end{align}
where $j = i + k$ and $\mathcal{I}_{s_i s_j}(\cdot)$ is the indicator function given by
\[
	\mathcal{I}_{s_i s_j}(t_i t_j) = \left\{ \begin{array}{cc}
		1 & \text{if }t_i = s_i \text{ and } t_j = s_j,\\
		0 & \text{otherwise.}
	\end{array} \right.
\]
Similarly, $w_{i,j}$ is set to either $1$ or the information content of the nucleotide pair
at $(i, j)$
%for all $i = 1, 2, \ldots, K$ and $j = 1, 2, \ldots, l-i$. 
given by
\begin{equation}\label{eq:IC_P}
	IC_{i, j} = 4 + \sum_{u,v \in \{\text{A, C, G, T}\}}f_{i, j}(u, v)\log_2\left[ f_{i, j}(u, v) \right],
\end{equation}
where $f_{i,j}(u, v)$ is the probability of observing letters $u$ and $v$ at positions $i$ and $j$,
respectively. We consider only pairs that are
at most 2 nucleotides apart ($K = 2$) according to the results reported in \cite{Osa04}.

\begin{figure}[!t]
\centering
\includegraphics[width = 0.4\textwidth]{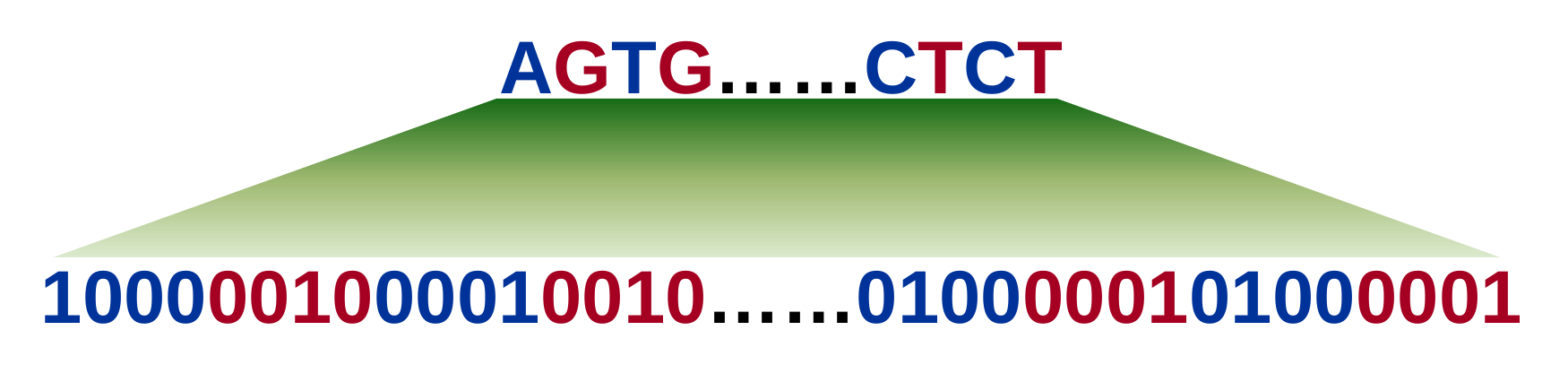}
% where an .eps filename suffix will be assumed under latex, 
% and a .pdf suffix will be assumed for pdflatex; or what has been declared
% via \DeclareGraphicsExtensions.
\caption{Illustration of embedding an $l$-mer in $\mathbb R^{4l}$ with $w_i = 1$ for $i = 1, 2, \ldots, l$.}
\label{fig:embedding}
\end{figure}

To facilitate similarity computation, an {\it l}-mer $s$ can be easily embedded in $\mathbb R^{4l}$ while preserving
the similarity measure in (\ref{eq:sim}) by the dot product between two vectors. That is, letter
$s_i$ is converted to 4 dummy variables --
$\sqrt{w_i}\mathcal{I}_{\text A}(s_i),  \sqrt{w_i}\mathcal{I}_{\text C}(s_i),
\sqrt{w_i}\mathcal{I}_{\text G}(s_i) \text{ and } \sqrt{w_i}\mathcal{I}_{\text T}(s_i)$
for $i=1, 2, \ldots, l$.
Fig.~\ref{fig:embedding} illustrates the transformation of an $l$-mer into a $4l$-element
vector when $w_i = 1$ for $i = 1, 2, \ldots, l$.
Similarly, an {\it l}-mer can be transformed into a 
%$(4l + 16(l-1 + l-2))$
$(36l - 48)$-element vector such that the similarity measure in (\ref{eq:sim2}) with $K=2$ is preserved,
where a pair of nucleotides is converted to 16 dummy variables.
Consequently, the similarity between two sequences $s$ and $t$,
can be computed by $\bm s\transpose\bm t$, where $\bm s$ and $\bm t$ denote sequences $s$ and $t$,
respectively, embedded in the Euclidean space. In the rest of the paper, we denote a sequence $s$
embedded in the Euclidean space by the same symbol in bold, i.e., $\bm s$.

Consider a set $S$ of $n_+$ binding sites of length $l$ for a TF. The centroid method scores
an {\it l}-mer $t$ by
\begin{align}
	\mathrm{Score}(t) 
	= \frac{1}{n_+}\sum_{s \in S} \bm s\transpose \bm t
	= \left(\frac{1}{n_+}\sum_{s \in S} \bm s \right)\transpose \bm t
	= \bm \mu_+\transpose \bm t\label{eq:centroid_score},
\end{align}
where $\bm \mu_+ = \frac{1}{n_+}\sum_{s \in S} \bm s$ is the centroid of the binding sites in $S$.

\begin{figure}[!t]
\centering
\includegraphics[width = 0.5\textwidth]{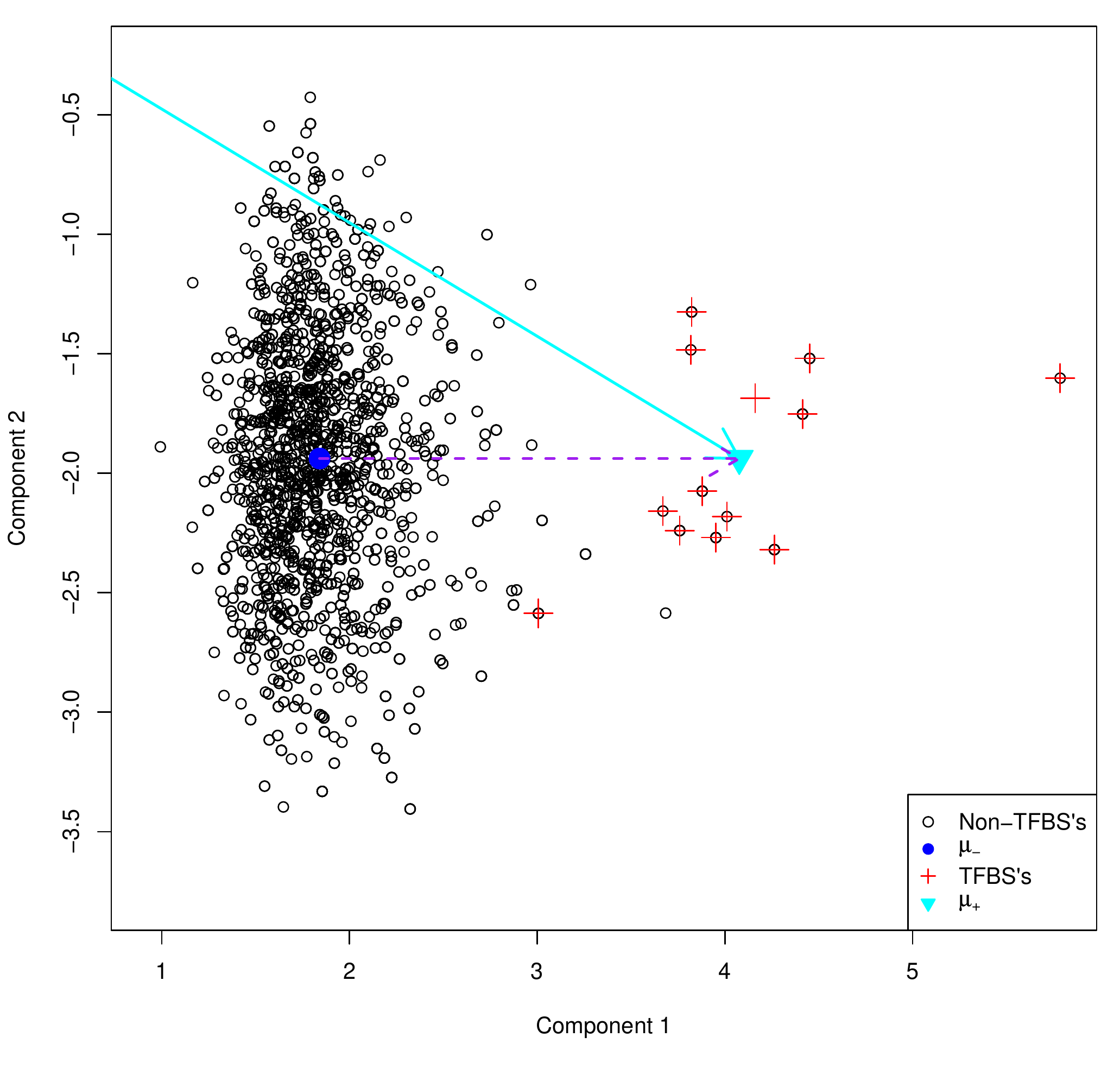}
\caption{Illustration of the 2-centroid method. The solid arrow denotes vector $\bm{\mu}_+$, while the dashed arrow
represents vector $\bm{\mu}_+ - \bm{\mu}_-$, pointing from $\bm{\mu}_-$ to $\bm{\mu}_+$.}
\label{fig:glpR_2centroids}
\end{figure}

Now, with a set $N$ of $n_-$ non-binding sites of length $l$ for the TF, a natural extension of
the centroid method scores an {\it l}-mer $t$ by
\begin{align}
	\mathrm{Score}(t) 
	&= \bm \mu_+\transpose \bm t - \frac{1}{n_-}\sum_{s \in N} \bm s\transpose \bm t
	= \bm \mu_+\transpose \bm t - \left(\frac{1}{n_-}\sum_{s \in N} \bm s \right)\transpose \bm t\notag\\
	&= (\bm \mu_+ - \bm \mu_-)\transpose \bm t\label{eq:centroid_neg_score},
\end{align}
where $\bm \mu_- = \frac{1}{n_-}\sum_{s \in N} \bm s$ is the centroid of the non-binding sites in $N$.
We refer to this method as the 2-centroid method in the rest of the paper since it employs the
centroids of the binding sites and the non-binding sites. Fig.~\ref{fig:glpR_2centroids} illustrates
the centroid and 2-centroid methods when non-TFBS's as well as TFBS's are available.
%, where the vector $\bm \mu_+$ is the solid arrow and $\bm \mu_+ - \bm \mu_-$ is the dashed arrow.
Alternatively, $\mathrm{Score}(t)$ in (\ref{eq:centroid_neg_score}) can be interpreted as follows: It
measures the average similarity of $t$ to all the binding sites, measures the average similarity of $t$
to all the non-binding sites and calculates the difference.

\begin{figure}[!tpb]
\centering
\includegraphics[width = 0.3\textwidth]{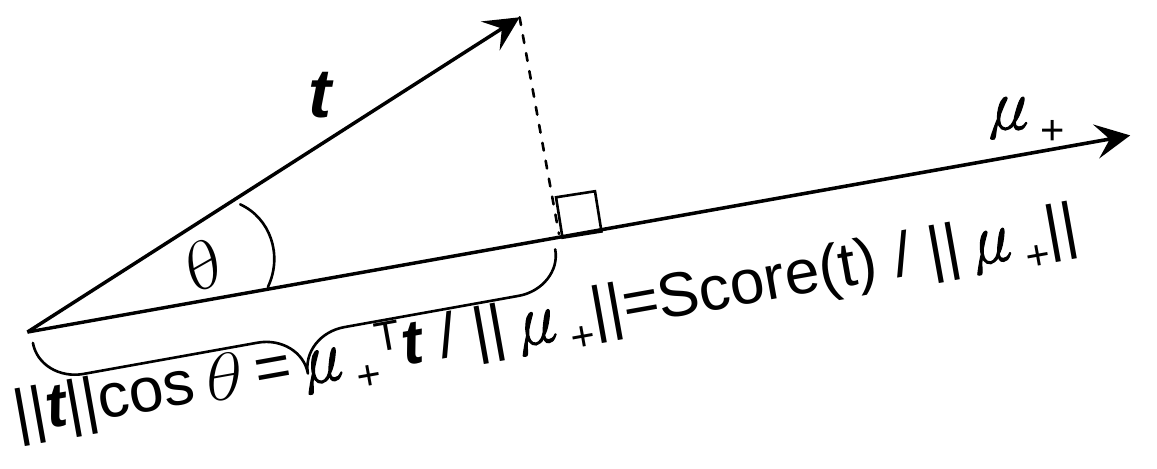}
\caption{The orthogonal projection of $\bm t$ onto $\bm \mu_+$ is equal to $\mathrm{Score}(t) / ||\bm \mu_+|| \propto \mathrm{Score}(t)$.}
\label{fig:projection}
\end{figure}

We note that $\mathrm{Score}(t)$ in (\ref{eq:centroid_score})
is proportional to $\mathrm{Score}(t) / ||\bm \mu_+||$ , where $||\bm \mu_+||$ is the length of $\bm \mu_+$.
% INTRODUCE ORTHOGONAL PROJECTION
Moreover, by virtue of the equality
\[
	\bm \mu_+\transpose \bm t = ||\bm \mu_+|| \,\, ||\bm t|| \cos \theta,
\]
we know $\mathrm{Score}(t) / ||\bm \mu_+||$ equals the orthogonal projection of $\bm t$ onto $\bm \mu_+$,
%$\frac{\bm \mu_+}{||\bm \mu_+||}\transpose \bm t$.
where $\theta$ is the angle formed by vectors $\bm \mu_+$ and $\bm t$ (see Fig.~\ref{fig:projection} for an illustration).
The computation of $\mathrm{Score}(t)$ is therefore equivalent to
computation of the orthogonal projection of $\bm t$ onto $\bm \mu_+$. Similarly, the computation of
$\mathrm{Score}(t)$ in (\ref{eq:centroid_neg_score}) is equivalent to computation of the orthogonal
projection of $\bm t$ onto $\bm \mu_+ - \bm \mu_-$.

\subsection{Optimal scoring function}\label{sec:ODV}

It can be seen that the scoring functions in (\ref{eq:centroid_score})~and~(\ref{eq:centroid_neg_score})
take the following form:
\begin{equation}\label{eq:scoring_function}
	\mathrm{Score}(t) = \bm \beta\transpose \bm t,
\end{equation}
where $\bm \beta = \bm \mu_+$ for the centroid method and $\bm \beta = \bm \mu_+ - \bm \mu_-$ for
the 2-centroid method. Therefore, an ``optimal'' $\bm \beta$ gives rise to an optimal scoring function
with the most discriminating power.

We describe a way of finding an optimal $\bm \beta$. Suppose that $|S| = n_+$ and $|N| = n_-$, that is,
there are $n_+$ binding sites and $n_-$ non-binding sites for a particular TF.
Let $S = \{t_{(1)}, t_{(2)}, \ldots, t_{(n_+)}\}$ and $N = \{t_{(n_+ + 1)}, t_{(n_+ + 2)}, \ldots, t_{(n)}\}$
, where $t_{(i)}$ denotes the $i^{\text{th}}$ {\it l}-mer in $S \cup N$
and $n = n_+ + n_-$. We find the optimal $\bm \beta$ by solving the following minimization
problem:
\begin{align}
	\min_{\bm \beta, b, \bm \xi}\,\, &\frac{1}{2}||\bm \beta||^2
	+ \frac{C}{n_+}\sum_{i=1}^{n_+}\xi_i + \frac{C}{n_-}\sum_{i=n_+ + 1}^{n}\xi_i\label{eq:obj_function}\\
	\text{subject to } & \frac{ \mathrm{Score}(t_{(i)}) } {||\bm \beta||} \geq
	\frac{ b + 1 - \xi_i }{||\bm \beta||} \text{ for } t_{(i)} \in S,\label{eq:constraint1}\\
	& \frac{ \mathrm{Score}(t_{(i)}) }{||\bm \beta||} 
	\leq \frac{ b - 1 + \xi_i }{||\bm \beta||} \text{ for } t_{(i)} \in N,\label{eq:constraint2}\\
	& \xi_i \geq 0 \,\,\forall i\label{eq:constraint3}.
\end{align}
The constraint in (\ref{eq:constraint1}) ensures that the projection of a TFBS $t_{(i)}$ onto the vector
$\bm \beta$, $\frac{\mathrm{Score}(t_{(i)})}{||\bm \beta||}$, exceeds the threshold
$\frac{b+1}{||\bm \beta||}$. On the other hand, the constraint in (\ref{eq:constraint2}) ensures that
the projection of a non-TFBS $t_{(i)}$ onto $\bm \beta$ stays below the threshold $\frac{b-1}{||\bm \beta||}$.
Flexibility is given to the thresholds by introducing $\xi_i$'s with cost captured by the last two terms in
(\ref{eq:obj_function}), where $C$ is a positive parameter. Finally, to clearly distinguish TFBS's from
non-TFBS's, the squared difference between the two thresholds ($\frac{b+1}{||\bm \beta||}$ and $\frac{b-1}{||\bm \beta||}$)
is made as large as possible. This amounts to maximizing $\left( \frac{2}{||\bm \beta||} \right)^2$ or,
equivalently, minimizing $\frac{1}{2}||\bm \beta||^2$, which is
the first term in (\ref{eq:obj_function}). We call this approach the optimal discriminating vector
(ODV) method.

\subsection{PSSM and ULPB}

We briefly describe the PSSM (position-specific scoring matrix) methods used in \cite{Osa04,Sal10}
and the ungapped likelihood under positional background method proposed by Salama and Stekel \cite{Sal10}.
%The PSSM method is also known as the position-specific probability matrix (PSPM) method.
%Given a set of known TFBS's of length $l$, we can estimate the probability of observing nucleotide
%$t$ at position $i$, denoted by $f_i(u)$, for $u \in \{\text{A, C, G, T}\}$ and $i = 1, 2, \ldots, l$.
Consider a specific TF with binding sites of length $l$.
The PSSM method used in \cite{Sal10} scores an $l$-mer $t$ by
\begin{equation}\label{eq:PSPM}
	\sum_{i=1}^l \log\left[ f_i(t_i) \right], %w_i,
\end{equation}
where
%the weight on the $i^{\text{th}}$ letter $w_i = 1$ for all $i$ and 
no pair of nucleotides was considered
for this model in \cite{Sal10}. We refer to this method as the position-specific \textit{probability} matrix
(PSPM) method to distinguish it from the PSSM used in \cite{Osa04}.

The PSSM method given in \cite{Osa04} takes into account background probabilities and scores an
$l$-mer by
\begin{equation}\label{eq:PSSM}
	\sum_{i=1}^l \log\left( \frac{f_i(t_i)}{f(t_i)} \right) w_i,
\end{equation}
where $f(u)$ is the probability of observing nucleotide $u \in \{$A, C, G, T$\}$.
When nucleotide pairs are considered, the score becomes
\begin{equation}\label{eq:PSSM_P}
	\sum_{i=1}^l w_i \log\left( \frac{f_i(t_i)}{f(t_i)} \right)
	+ \sum_{k = 1}^K \sum_{i=1}^{l-k} w_{i,j} \log\left( \frac{f_{i,j}(t_i, t_j)}{f_k(t_i, t_j)} \right),
\end{equation}
where $j = i + k$, $K = 2$ and $f_k(u, v)$ is the background probability of observing
letters $u$ and $v$ separated by $k-1$ arbitrary letters in between. For this method,
we estimate the background probabilities using only the TFBS sequences as in \cite{Osa04}.

The ULPB models a TFBS by a first-order Markov chain and models the background by another
first-order Markov chain.
The former depends on position-specific transition probability $f_i(v|u)$, which
gives the probability of observing $v$ at the $(i+1)^{\text{th}}$ position given
$u$ has been seen at position $i$, where $u, v \in \{\text{A, C, G, T}\}$
and $i = 1, 2, \ldots, l-1$.
The latter depends on background transition probability $f(v|u)$, the probability of observing
$v$ given $u$ has been observed at the previous position, where $u, v \in \{\text{A, C, G, T}\}$.
For this method, the background transition probabilities are estimated using the entire genome of a species.
The ULPB method scores an $l$-mer by
\begin{equation}\label{eq:ULPB}
	\log f_1(t_1) + \sum_{i = 1}^{l-1} \log \left( \frac{f_i(t_{i+1}|t_i)}{f(t_{i+1}|t_i)} \right).
\end{equation}
Although Salama and Stekel \cite{Sal10} did not consider background probability in the first term of (\ref{eq:ULPB}),
the score is approximately the log-likelihood ratio of the two Markov chains.

% MOVE TO DISCUSSION:
% \subsection{Connection between ODV and PSSM/ULPB}\label{sec:connection}

\section{Results}\label{sec:results}

In this section, we show results of experiments conducted on the two data sets introduced in Section~\ref{sec:dataset}.
Results on the first data set are presented in Section~\ref{sec:LOO} through Section~\ref{sec:LOO2},
while results on the second set are summarized in Sections~\ref{sec:ULPB_comp}.
%~and~\ref{sec:visualizeODV}.

\subsection{Leave-one-out cross-validation}\label{sec:LOO}

We conducted LOO CV experiments on the data set introduced in the previous section. To allow comparison of our results
to those obtained by Osada \textit{et al.} \cite{Osa04}, we closely followed the steps described in
\cite{Osa04}. We briefly describe the LOO CV procedure adopted in \cite{Osa04} since only the TFBS's are
left out in the process.

Consider a TF with $n_+$ TFBS's of length $l$ with flanking regions on both sides. A set of negative
examples, $N_{\text{test}}$, called the \emph{test negatives} is constructed from the TFBS's of the
other 34 TF's as in \cite{Osa04}. Another set of negative examples, $N_{\text{train}}$, called the
\emph{training negatives} is collected from sequences embedding the $n_+$ binding sites. It comprises
all the {\it l}-mers except for the TFBS's and two neighboring {\it l}-mers of each TFBS.

At each iteration of LOO CV, one of the $n_+$ TFBS's called the \emph{test TFBS} is left out.
The rest of the TFBS's are therefore called the \emph{training TFBS's}. A scoring function is then
obtained using the training TFBS's and 5\% of non-TFBS's randomly sampled from the training negatives.
The test TFBS along with the non-TFBS's in $N_{\text{test}}$ are then scored by the scoring function.
To score a test sequence, both the forward and reverse strands are scored and, in case the test sequence
is longer or shorter than $l$, the {\it l}-mer producing the highest score is used. The rank of the test TFBS
is then recorded and the average rank over the CV process is computed, where the rank of a TFBS $t$ is
defined as $1 + |\{s \in N_{\text{test}} | \mathrm{Score}(s) \geq \mathrm{Score}(t)\}|$.

\begin{figure*}[!t]
\centering
\subfloat[{ }]{\label{fig:boxplot}
\includegraphics[width = 0.5\textwidth]{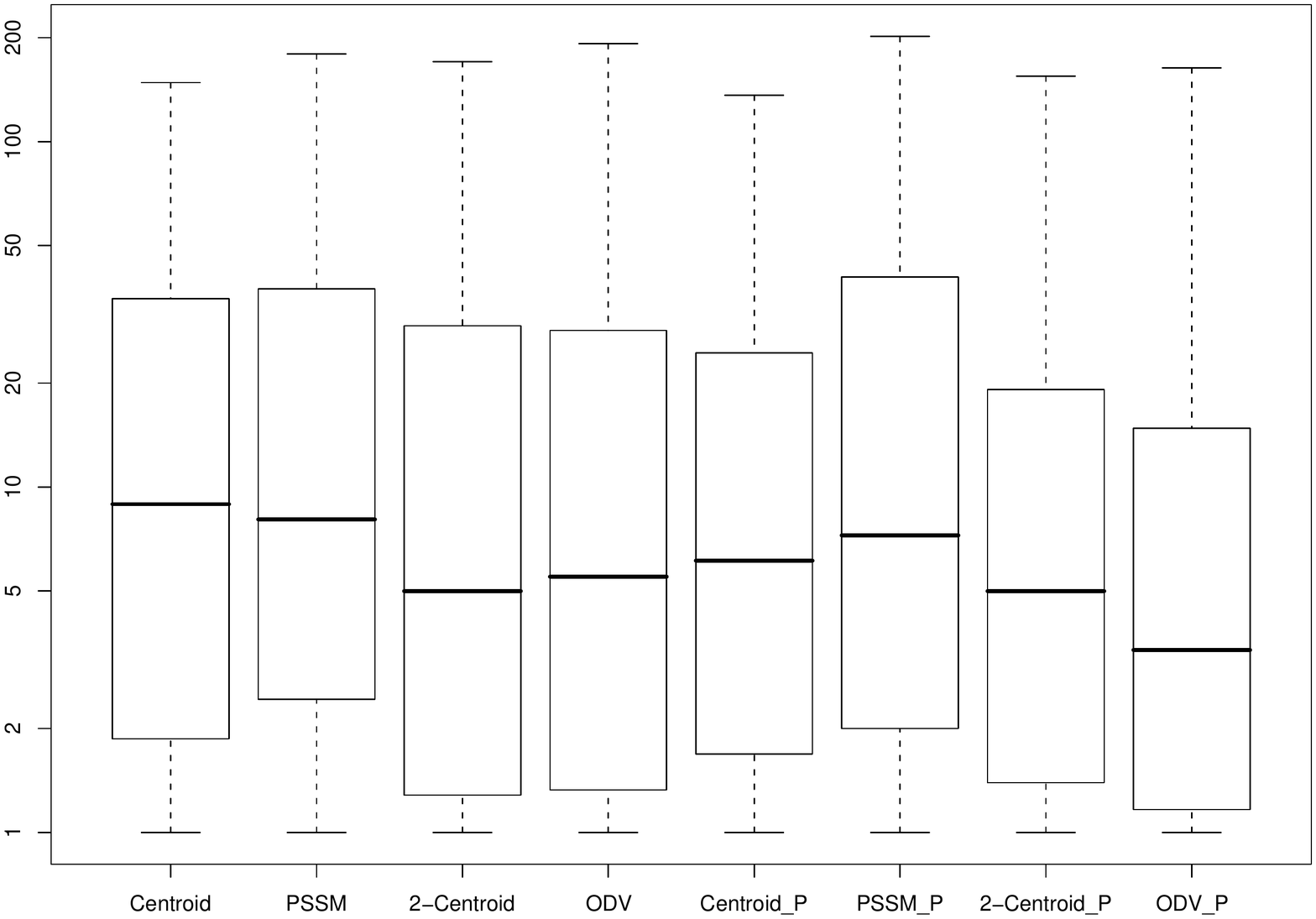}
}
\subfloat[{ }]{\label{fig:boxplot_IC}
\includegraphics[width = 0.5\textwidth]{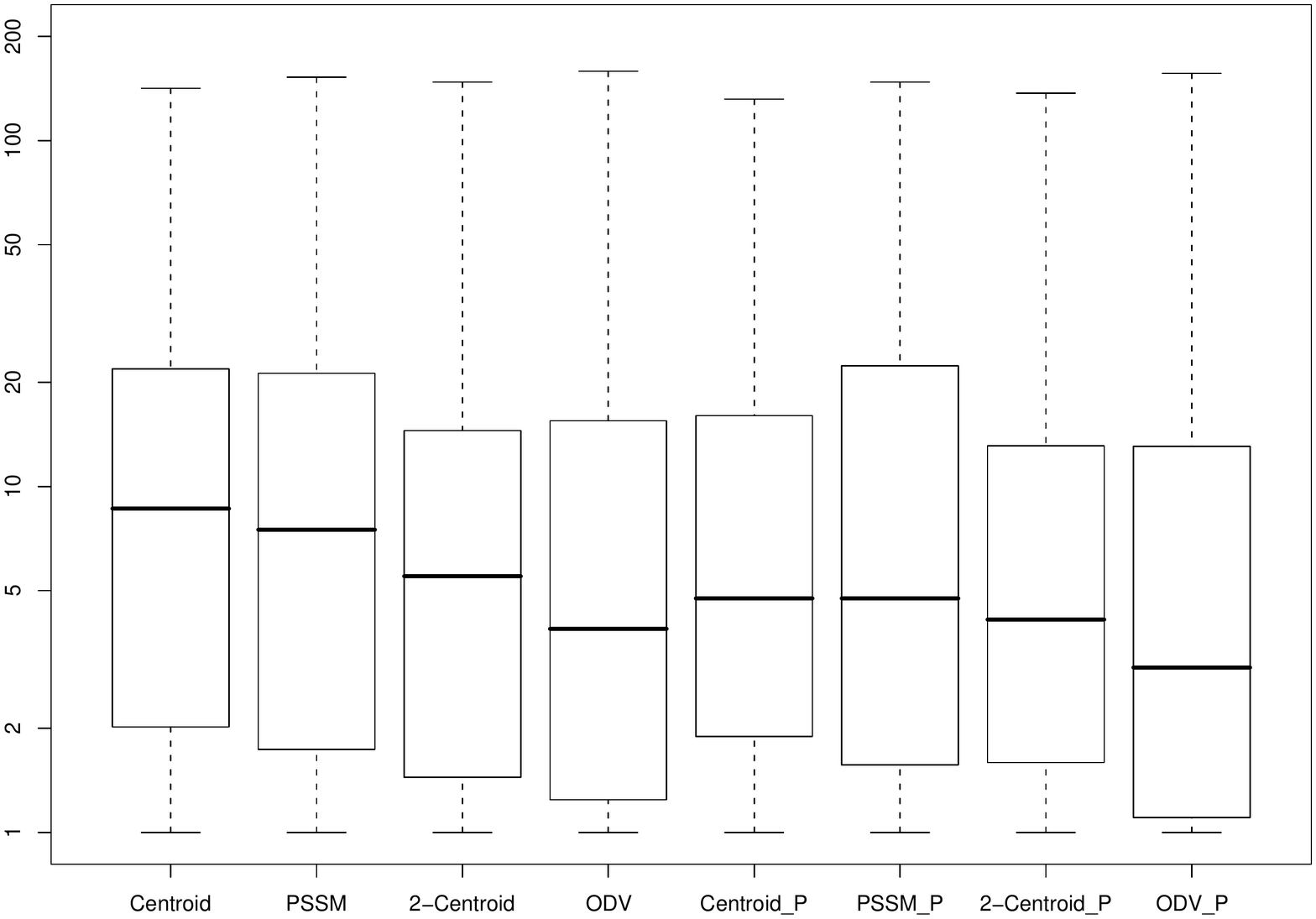}
}
\caption{
Box plots of average ranks of the 35 TF's.
A box contains TF's with ranks falling between
the $25^{\text{th}}$ and $75^{\text{th}}$ percentiles, while the median is marked by the horizontal
bar in it. The ends of the whiskers mark the minimum and maximum of average ranks of all the TF's.
A suffix ``\_P'' in name means that the similarity measure given in (\ref{eq:sim2}) or the score
in (\ref{eq:PSSM_P}) is used. (a) Each nucleotide or nucleotide pair is given the same weight.
(b) Each nucleotide or nucleotide pair is weighted by its information content.}
\label{fig:boxplot_LOOCV}
\end{figure*}

In this study, the weight on nucleotide $i$, $w_i$, is set to either 1 or its information content given in
(\ref{eq:IC}). Similarly, the weight on a nucleotide pair, $w_{i,j}$ is set to either 1 or its information
content defined in (\ref{eq:IC_P}).
Fig.~\ref{fig:boxplot_LOOCV} shows the LOO CV results as box plots without and
with information content, respectively.
The best run over 10 runs is listed for a method utilizing the training negatives.
%A box contains TF's with ranks falling between
%the $25^{\text{th}}$ and $75^{\text{th}}$ percentiles, while the median is marked by the horizontal
%bar in it. The ends of the whiskers mark the minimum and maximum of the 35 average ranks.
%A suffix ``\_P'' in name means that the similarity measure given in (\ref{eq:sim2}) or the score
%in (\ref{eq:PSSM_P}) is used.
Results on the centroid and PSSM methods reported in \cite{Osa04} were faithfully reproduced here.
Moreover, from the box plots, we can see that methods utilizing negative examples perform better than
methods considering only positive examples.

%\begin{figure}[!t]
%\centering
%\includegraphics[width = 0.5\textwidth]{boxplot.pdf}
%\caption{Box plots of average ranks of the 35 TF's.
%Each nucleotide or nucleotide pair is given the same weight.
%A box contains TF's with ranks falling between
%the $25^{\text{th}}$ and $75^{\text{th}}$ percentiles, while the median is marked by the horizontal
%bar in it. The ends of the whiskers mark the minimum and maximum of average ranks of all the TF's.
%A suffix ``\_P'' in name means that the similarity measure given in (\ref{eq:sim2}) or the score
%in (\ref{eq:PSSM_P}) is used.
%}
%\label{fig:boxplot}
%\end{figure}

%\begin{figure}[!t]
%\centering
%\includegraphics[width = 0.5\textwidth]{boxplot_IC.pdf}
%\caption{Box plots of average ranks of the 35 TF's. Each nucleotide or nucleotide pair is
%weighted by its information content.
%A box contains TF's with ranks falling between
%the $25^{\text{th}}$ and $75^{\text{th}}$ percentiles, while the median is marked by the horizontal
%bar in it. The ends of the whiskers mark the minimum and maximum of average ranks of all the TF's.
%A suffix ``\_P'' in name means that the similarity measure given in (\ref{eq:sim2}) or the score
%in (\ref{eq:PSSM_P}) is used.
%}
%\label{fig:boxplot_IC}
%\end{figure}

\begin{figure*}[!t]
\centering
\subfloat[{ }]{\label{fig:Wilcox}
\includegraphics[width = 0.5\textwidth]{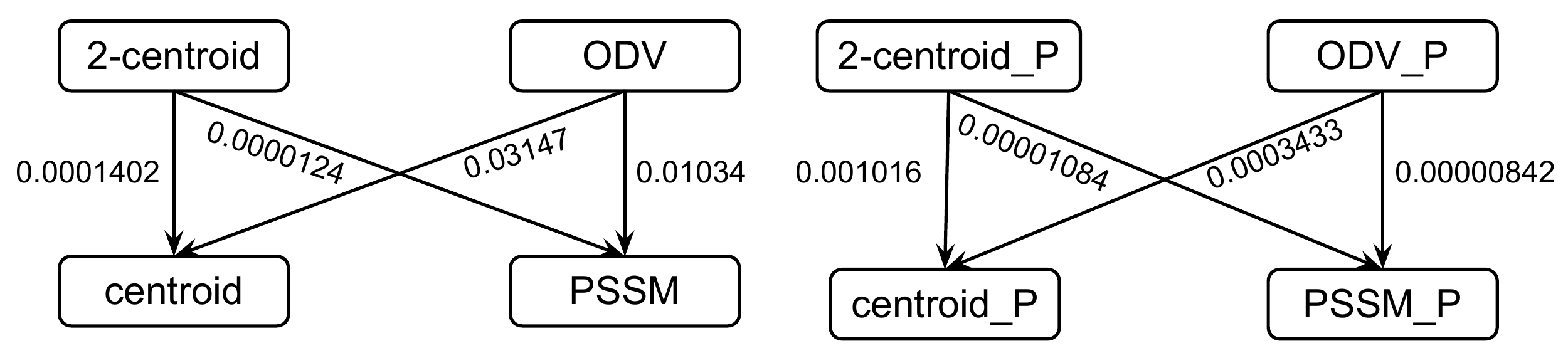}
}
\subfloat[{ }]{\label{fig:Wilcox_IC}
\includegraphics[width = 0.5\textwidth]{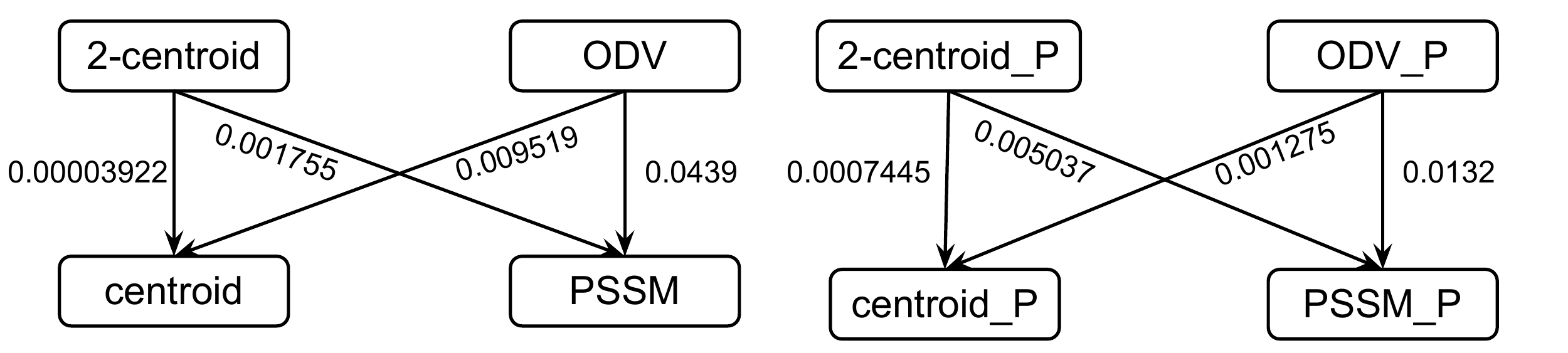}
}
\caption{
Results of Wilcoxon signed-rank tests on 4 pairs of methods (a) without IC and (b) with IC. Arrows along with $p$-values
point from the superior method to the inferior one.}
\label{fig:Wilcox_LOOCV}
\end{figure*}

To test whether the 2-centroid and ODV methods produced lower average ranks than the centroid and PSSM
methods, we adopted the testing procedure used in \cite{Osa04}.
The Wilcoxon signed-rank test \cite{Wil45} was performed on four pairs of methods.
They are (centroid, 2-centroid), (PSSM, 2-centroid), (centroid, ODV) and (PSSM, ODV). Multiple testing
was corrected by the Holm-Bonferroni method \cite{Hol79}.
The testing was done for each of the 4 similarity measures, i.e., $\mathrm{Sim}$ and $\mathrm{Sim2}$
in (\ref{eq:sim})~and~(\ref{eq:sim2}), respectively, with or without weighting by information content.
Results showed that, at 5\% significance level, the following relationships can be justified for
each similarity measure:
2-centroid $\rightarrow$ centroid, 2-centroid $\rightarrow$ PSSM,
ODV $\rightarrow$ centroid and ODV $\rightarrow$ PSSM,
where ``$\rightarrow$'' denotes ``has a lower average rank than''.
Fig.~\ref{fig:Wilcox}~and~\ref{fig:Wilcox_IC} show the $p$-values of the tests on 4 pairs of methods
without IC and with IC, respectively.

%\begin{figure}[!t]
%\centering
%\includegraphics[width = 0.5\textwidth]{Wilcox_Test.pdf}
%\caption{Results of Wilcoxon signed-rank tests on 4 pairs of methods without IC. Arrows along with $p$-values
%point from the superior method to the inferior one.}
%\label{fig:Wilcox}
%\end{figure}

%\begin{figure}[!t]
%\centering
%\includegraphics[width = 0.5\textwidth]{Wilcox_Test_IC.pdf}
%\caption{Results of Wilcoxon signed-rank tests on 4 pairs of methods with IC. Arrows along with $p$-values
%point from the superior method to the inferior one.}
%\label{fig:Wilcox_IC}
%\end{figure}

\subsection{The 2-centroid method with a novel similarity measure}

Do and Wang \cite{DoH09} proposed a novel distance measure by first transforming a
sequence of length $l$ into an $(l-1)$-element vector. To measure the distance between
two sequences $s$ and $t$, $t$ can be shifted to the left or to the right (with penalty)
to find the best alignment between $s$ and $t$. Since shifting is implicitly done in
scoring a non-binding site in our CV experiments, we use the distance measure without
considering shifting:
\begin{equation}\label{eq:dist}
	\mathrm{Dist}(\bm s, \bm t) = \sum_{i=1}^{l-1} |s_i - t_i|,
\end{equation}
where $\bm s = \begin{pmatrix}s_1 & s_2 & \ldots & s_{l-1}\end{pmatrix}$ and
$\bm t = \begin{pmatrix}t_1 & t_2 & \ldots & t_{l-1}\end{pmatrix}$ are the sequences $s$
and $t$ embedded in $\mathbb R^{l-1}$, respectively. One can see that this is essentially the
Manhattan distance between $\bm s$ and $\bm t$. To compute the similarity between $s$ and
$t$, we take the negative distance as the similarity.

\begin{figure*}[!t]
\centering
\includegraphics[width=\textwidth]{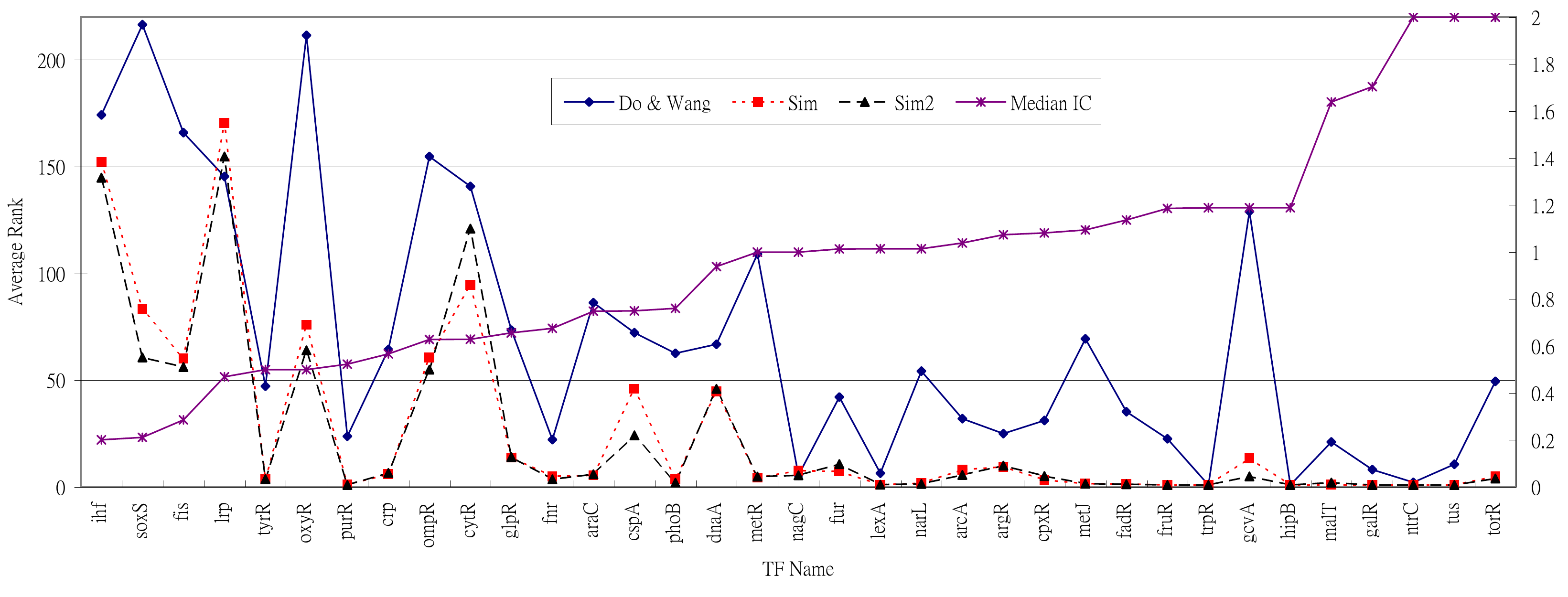}
\caption{Comparison of three similarity measures using the 2-centroid method.}
\label{fig:comp}
\end{figure*}

This similarity measure is then used along with our 2-centroid method. Fig.~\ref{fig:comp} compares
the performance of the similarity measures $\mathrm{Sim}$ in (\ref{eq:sim}) ($w_i = 1, \,\,\forall i$)
and $\mathrm{Sim2}$ in (\ref{eq:sim2}) ($w_i = 1, \,\,\forall i$ and $w_{i,j} = 1, \,\,\forall i, j$)
to the one proposed in \cite{DoH09}. The TF's are ordered by their median information content
across the $l$ nucleotides, i.e., the median of $\{IC_i| i = 1, 2, \ldots, l\}$. A general trend can be
observed, that is, the performance of a method improves as the median information content increases.
Looking at individual TF's, we can see that the similarity measure by Do and Wang %\cite{DoH09}
gave the lowest average rank on TF lrp, performed equally well on TF's hipB and trpR, but produced
the highest average ranks on all the other TF's.

\subsection{Yet another LOO CV}\label{sec:LOO2}

Two different sets of negative examples were used in the LOO CV experiments presented above
since no prior knowledge of the test negatives was assumed.
We now show that, with the knowledge of non-binding sites, a small representative set of negative
examples can be found by a slightly different LOO CV procedure. To avoid ambiguity, we constantly
refer to sets defined in Section~\ref{sec:LOO}.
%the previous LOO CV procedure.

Consider a particular TF with $n_+$ known TFBS's of length $l$. Suppose that the goal is to search
for sites to which this TF binds but avoid known binding sites of other TF's. That is, the
binding sites of the other 34 TF's are assumed known. We first randomly sample a representative set
of $10n_+$ {\it l}-mers, $N_{\text{rep}}$, from $N_{\text{test}}$ since $10n_+ \approx 0.05 |N_{\text{train}}|$.
For each iteration of LOO CV, the test TFBS is left out. A scoring function is
obtained using the $n_+ - 1$ training TFBS's and $N_{\text{rep}}$. The rank of the test TFBS is then
calculated based on its score and the scores of the non-TFBS's in $N_{\text{test}}$. The average rank
of this TF is computed at the end of the LOO CV procedure. A good representative set of $10n_+$
negative examples can be found by repeating this LOO CV procedure multiple times.

\begin{figure}[!t]
\centering
\includegraphics[width = 0.5\textwidth]{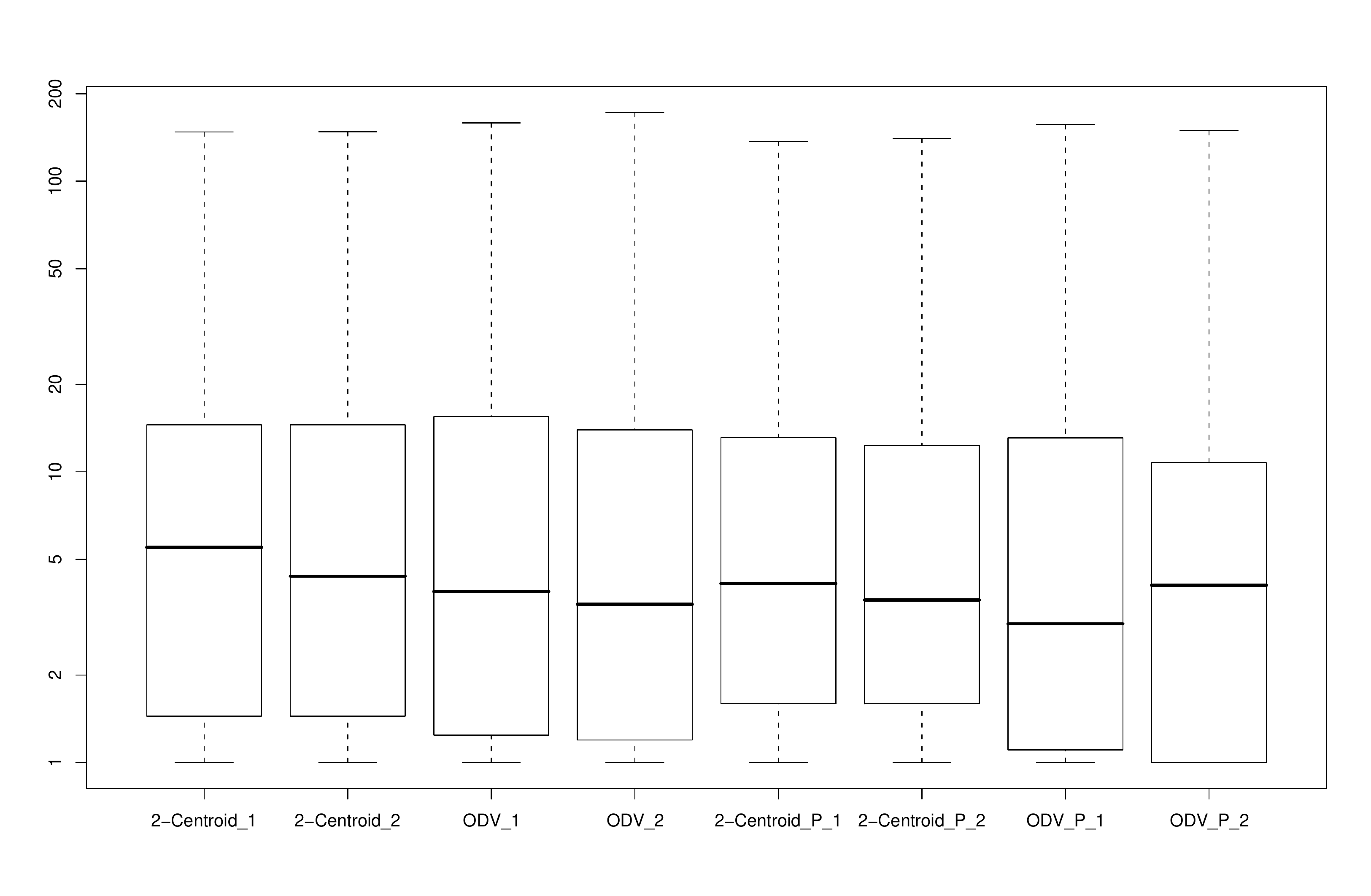}
\caption{Box plots of average ranks of the 35 TF's.
Each nucleotide or nucleotide pair is weighted by its information content.}
\label{fig:boxplot_IC2}
\end{figure}

We sampled a representative set of negative examples for each TF by repeating the LOO CV procedure 32
times. Fig.~\ref{fig:boxplot_IC2} compares average ranks resulted from the LOO CV procedure described
in this section to those obtained in the first set of LOO CV experiments.
Results of the first LOO CV procedure are marked with suffix ``\_1'', while those of the LOO CV
experiments described in this section are marked with suffix ``\_2''. As expected, the average ranks obtained
from the second set of LOO CV experiments are lower or comparable to those obtained from the first set.
Looking at the medians of ODV\_P\_1 and ODV\_P\_2, it may appear that ODV\_P\_2 performed worse than
ODV\_P\_1. However, a statistical test \cite{Wil45} indicates that overall ODV\_P\_2 has lower average ranks
than ODV\_P\_1 ($p$-value: 0.06975).

\subsection{ULPB versus other methods}\label{sec:ULPB_comp}

Since the ungapped likelihood under positional background method was evaluated by Salama and Stekel \cite{Sal10} on a data set
collected from RegulonDB, we conducted LOO CV experiments using the second data set described in Section~\ref{sec:dataset}.
The methods compared to ULPB include the position-specific probability matrix (PSPM) method, the position-specific scoring
matrix method with nucleotide pairs (PSSM\_P), the 2-centroid method with nucleotide pairs (2-centroid\_P)
and the optimal discriminating vector with nucleotide pairs (ODV\_P).
PSPM was chosen because it was one of the methods compared in \cite{Sal10}.
PSSM\_P was included because it does not require non-TFBS's and it is similar to ULPB in that nucleotide pairs
are considered. ODV\_P and 2-centroid\_P were compared because they employ non-TFBS's explicitly.
Information content was not used in all the methods compared in this section.

\begin{figure}[!t]
\centering
\includegraphics[width = 0.4\textwidth]{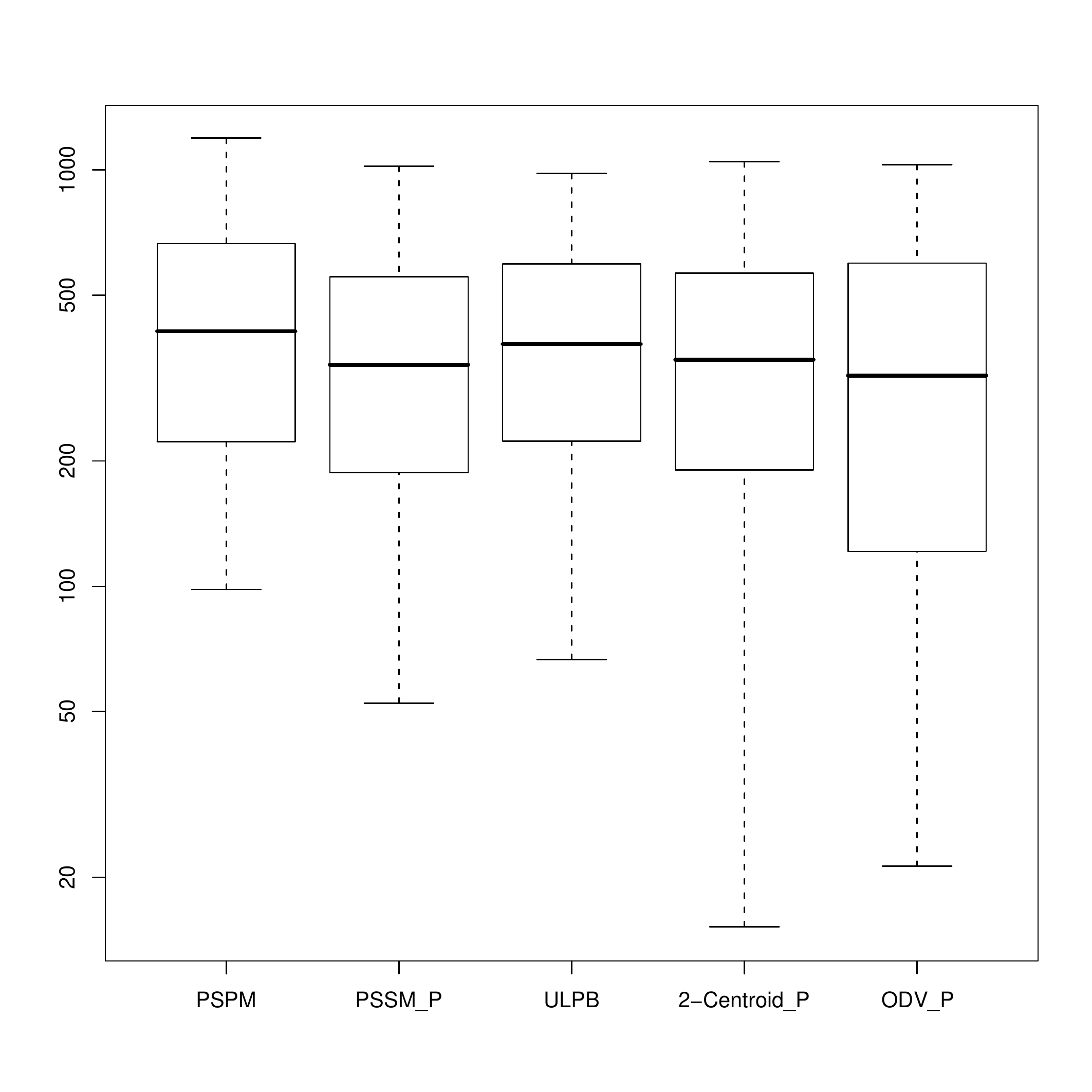}
\caption{Box plots of average ranks of the 26 TF's in the second data set.}
\label{fig:boxplot_Sal10}
\end{figure}

\begin{figure*}[!t]
\centering
\includegraphics[width=\textwidth]{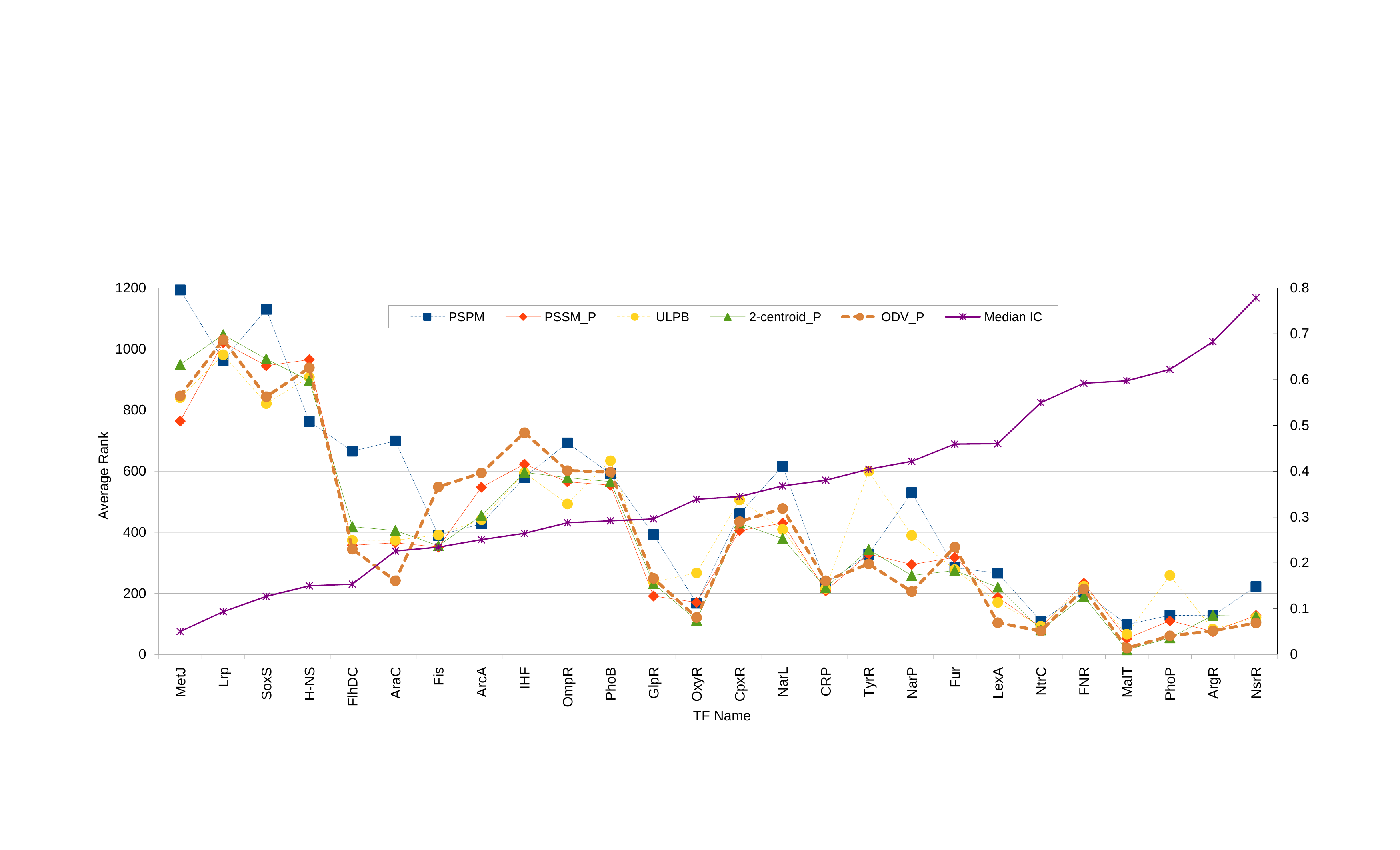}
\caption{Comparison of the PSPM, PSSM\_P, ULPB, 2-centroid\_P and ODV\_P methods using the second data set.}
\label{fig:comp_Sal10}
\end{figure*}

The methods were evaluated under the same LOO CV framework described in Section~\ref{sec:LOO}.
Overall performance of the compared methods is summarized in Fig.~\ref{fig:boxplot_Sal10}.
The box plots show that overall PSPM gave the highest average ranks, which is consistent with the results reported in
\cite{Sal10} that ULPB performed better than PSPM. In terms of median marked by the horizontal bar inside a box,
ULPB appears to be worse than PSSM\_P, 2-centroid\_P and ODV\_P. Fig.~\ref{fig:comp_Sal10} shows performance of
the 4 methods on individual TF's. We can see that PSSM\_P performed better than ULPB on 15 out of 26 TF's
and 2-centroid\_P/ODV\_P performed better than ULPB on 14 out of 26 TF's. To gauge the significance of these observations,
statistical tests \cite{Wil45} were performed on all the 6 pairs of methods. The results however only support that
2-centroid\_P outperformed PSPM ($p$-value: 0.000722), ODV\_P outperformed PSPM ($p$-value: 0.03344)
and PSSM\_P outperformed PSPM ($p$-value: 0.006476).
The $p$-values of the other tests are all greater than 5\%, the usual significance cut-off.
Similar to Fig.~\ref{fig:comp}, the relation between performance and median information content can be
observed as well.

\section{Discussion}\label{sec:discussion}

\subsection{No best method for all TF's}

We have shown in the previous section that overall methods utilizing negative examples perform
better than methods using only positive examples.
One may be tempted to identify the method that gives the lowest average rank for all the TF's.
From the results of our LOO CV experiments, however, we found that there's no combination of
method and similarity measure that is optimal for all the TF's in the data sets. That is,
introducing pairs of nucleotide in similarity computation or incorporating non-binding sites
lowers the average ranks for most of the TF's but increases the average ranks for a few of them.
Fig.~\ref{fig:comp} serves as an example. It shows that the similarity measure proposed by
\cite{DoH09} gives the highest average ranks for most of the TF's but is the best
one among the three measures for TF lrp when the 2-centroid method is used. It also shows that
$\mathrm{Sim2}$ yields lower average ranks than $\mathrm{Sim}$ except for a few TF's such as
cytR and fur when used along with the 2-centroid method.
Therefore, instead of finding the combination of similarity measure and method that is optimal
for all the TF's. It is more reasonable and practical to search for the best combination of
similarity measure and method for a particular TF of interest, which can be achieved by
CV experiments.

\subsection{Complexity of transcription factor binding sites}

\begin{figure}[!t]
\centering
\includegraphics[width = 0.5\textwidth]{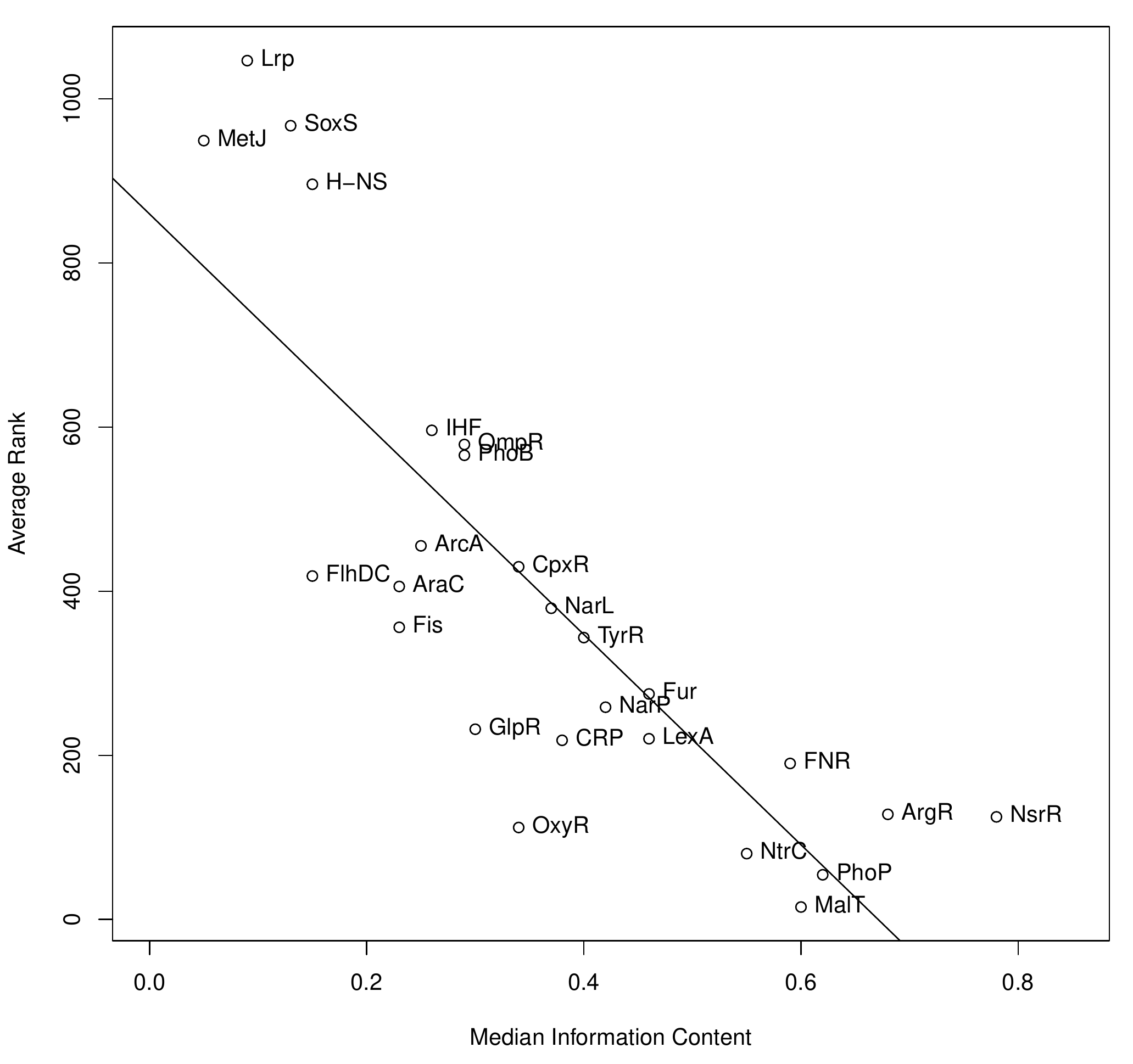}
\caption{Linear relationship between average rank and median information content. The average ranks
were obtained by running 2-centroid\_P without weighting by information content on the second data set.}
\label{fig:regression}
\end{figure}

Results presented in Fig.~\ref{fig:comp}~and~\ref{fig:comp_Sal10} indicate correlation between
the ``complexity'' of a TF and its median information content across nucleotides.
Therefore, we attempted to establish the relationship between average rank and three factors:
the length, number of known TFBS's and median information content.
The average ranks on the second data set produced by 2-centroid\_P in Fig.~\ref{fig:comp_Sal10} were linearly
regressed \cite{Rav01} on the three factors. Aside from the intercept, only the median information content
was found significant ($p$-value: $2.89 \times 10^{-7}$). A simple linear regression was then performed to
obtain the linear relationship between average rank and median information content. Fig.~\ref{fig:regression}
shows a scatter plot of average rank versus median information content for the 26 TF's in the second data set.
The straight line represents the relationship between average rank and median information content
found by simple linear regression. The median information content can be viewed as a measure of conservedness
of binding sites of a TF. This reasonably implies that the binding sites of a TF are easier to predict when they
are more conserved.

\subsection{Properties of Investigated Methods}\label{sec:pairwise_comp}

\begin{figure}[!t]
\centering
\includegraphics[width = 0.5\textwidth]{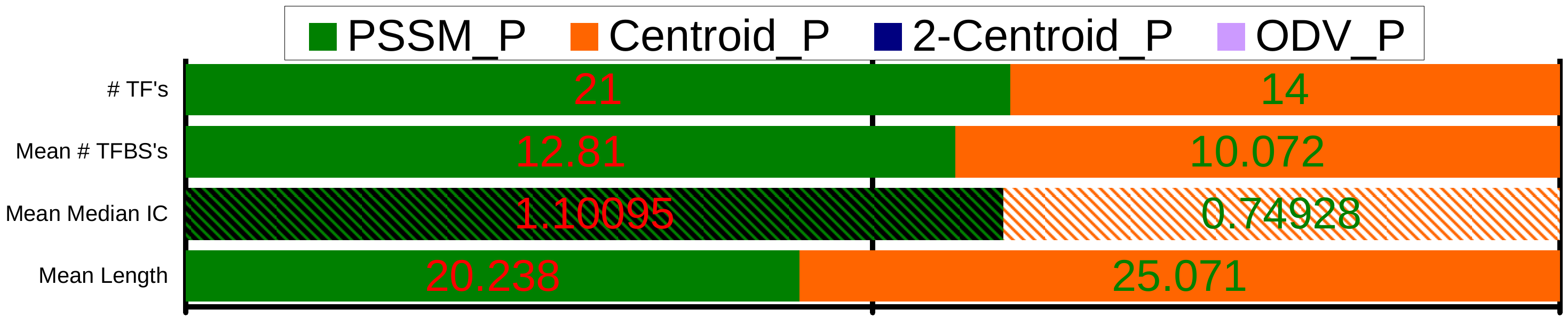}
\includegraphics[width = 0.5\textwidth]{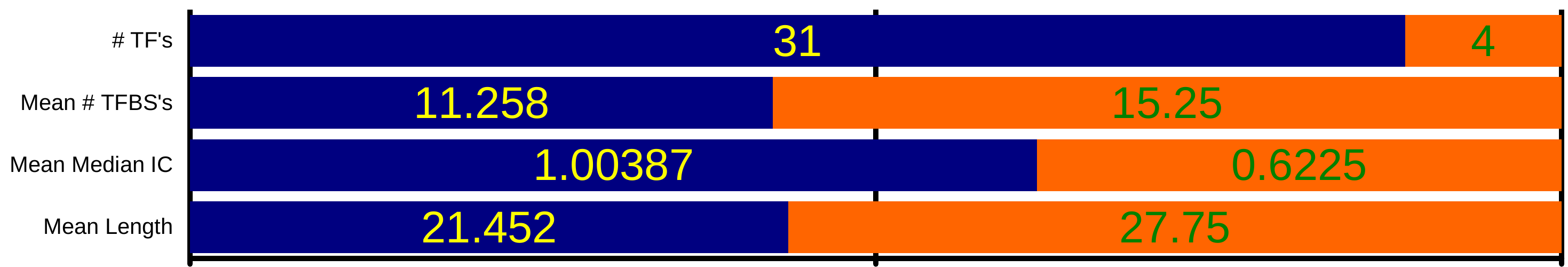}
\includegraphics[width = 0.5\textwidth]{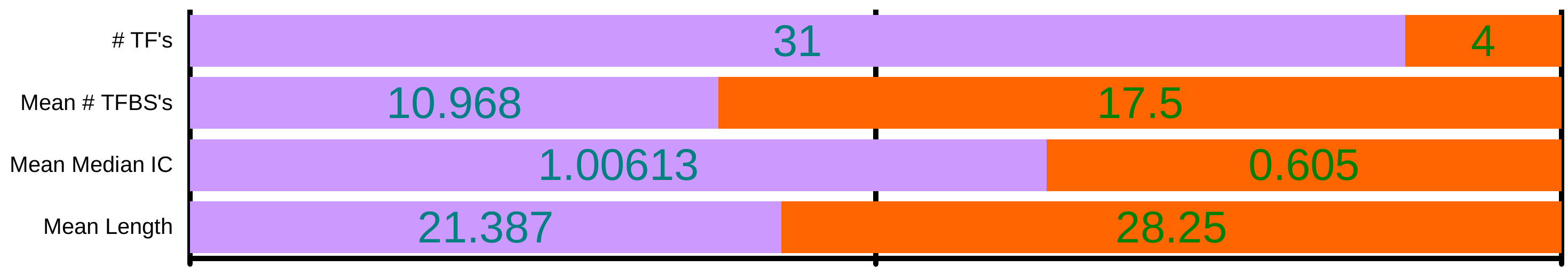}
\includegraphics[width = 0.5\textwidth]{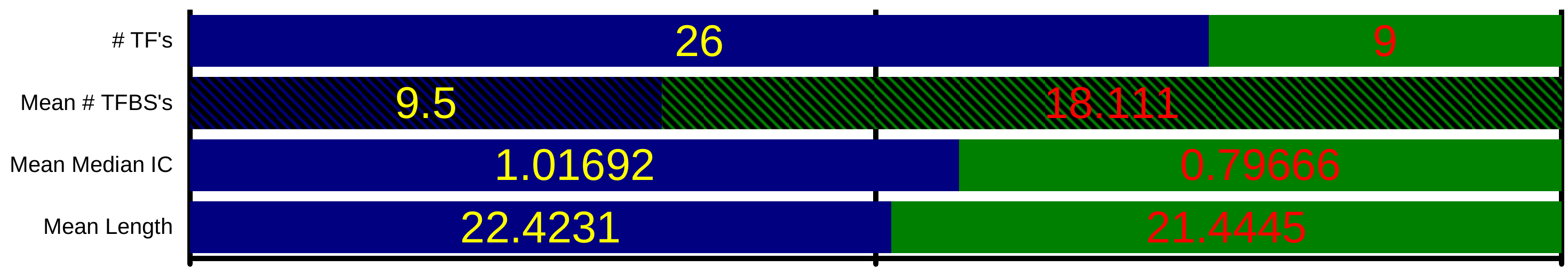}
\includegraphics[width = 0.5\textwidth]{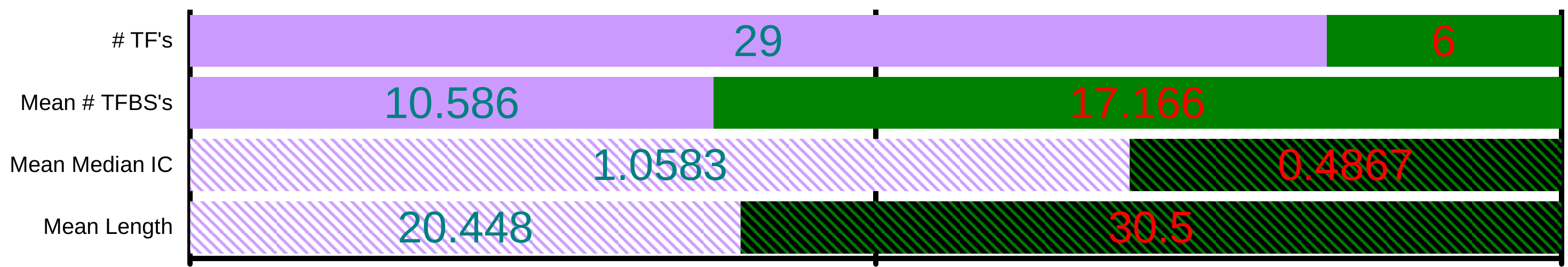}
\includegraphics[width = 0.5\textwidth]{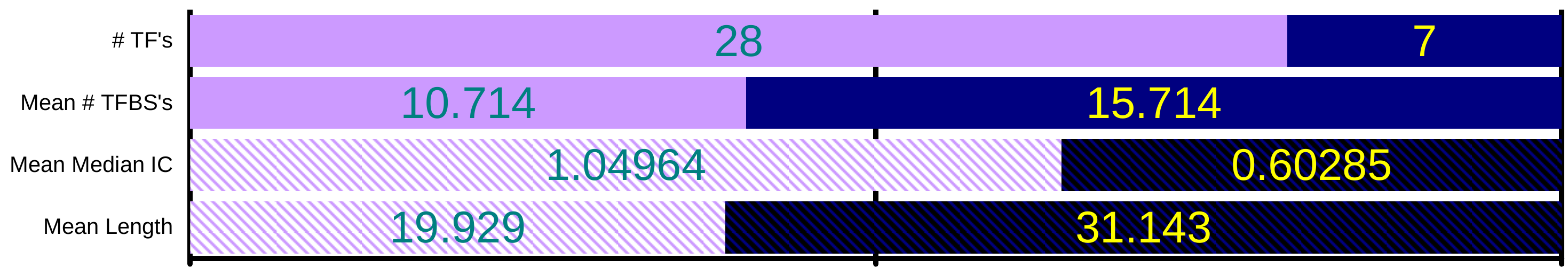}
\caption{Pair-wise comparisons of centroid\_P, PSSM\_P, 2-centroid\_P and ODV\_P with information content
on the first data set of 35 TF's.
Three factors except for \# TF's are tested for statistical significance. Significant factors are marked
by striped bars.}
\label{fig:pairwise_comp}
\end{figure}

To reveal properties of methods, we performed pair-wise comparisons on some of the methods investigated in this work.
Fig.~\ref{fig:pairwise_comp} shows the pair-wise comparisons of centroid\_P, PSSM\_P, 2-centroid\_P
and ODV\_P with information content on the first data set.
For each pair of methods, the 35 TF's were divided into two groups depending on the performance of the methods.
We then looked for statistical difference between the two groups in terms of three factors, that is,
the number of known TFBS's, the median IC and the length of binding sites.
The comparison between centroid\_P and PSSM\_P indicates that PSSM\_P performs better than centroid\_P on 21 TF's,
i.e., there are 21 TF's in one group and 14 TF's in the other. Moreover, when PSSM\_P performs better,
the median IC of a TF is on average 1.10095, which is significantly ($p$-value $< 5\%$) greater than
0.74928, the average median IC of a TF when centroid\_P performs better. Similar interpretations
lead to additional comments as follows. 2-centroid\_P requires significantly less known TFBS's than PSSM\_P.
ODV\_P performs better than PSSM\_P or 2-centroid\_P when a TF has higher median IC and shorter binding sites.

\begin{figure}[!t]
\centering
\includegraphics[width = 0.5\textwidth]{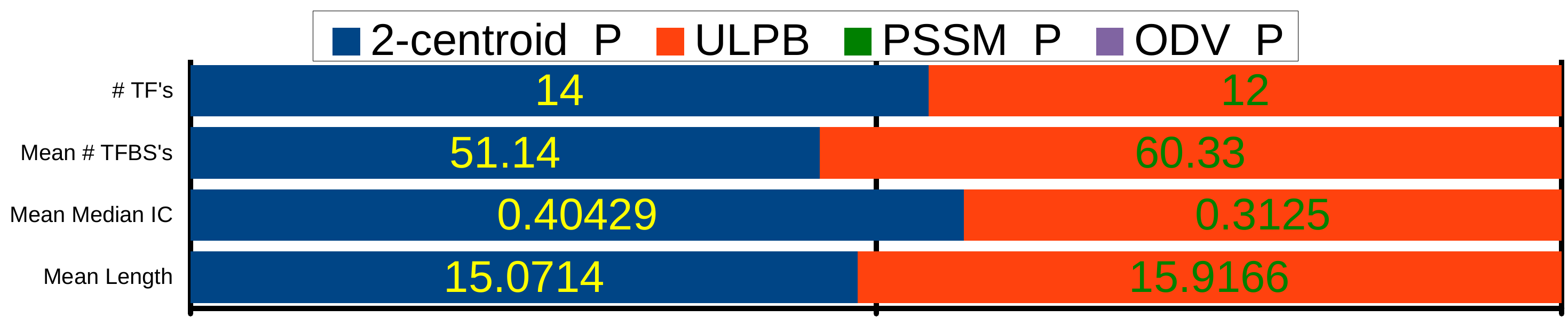}
\includegraphics[width = 0.5\textwidth]{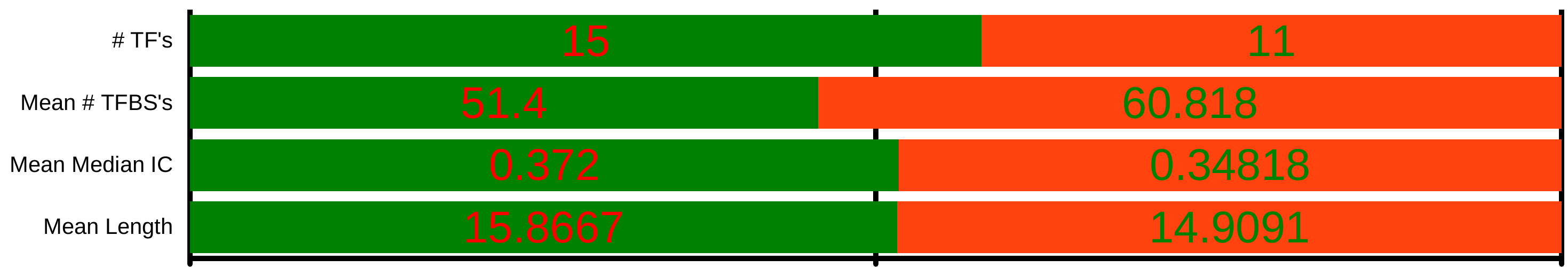}
\includegraphics[width = 0.5\textwidth]{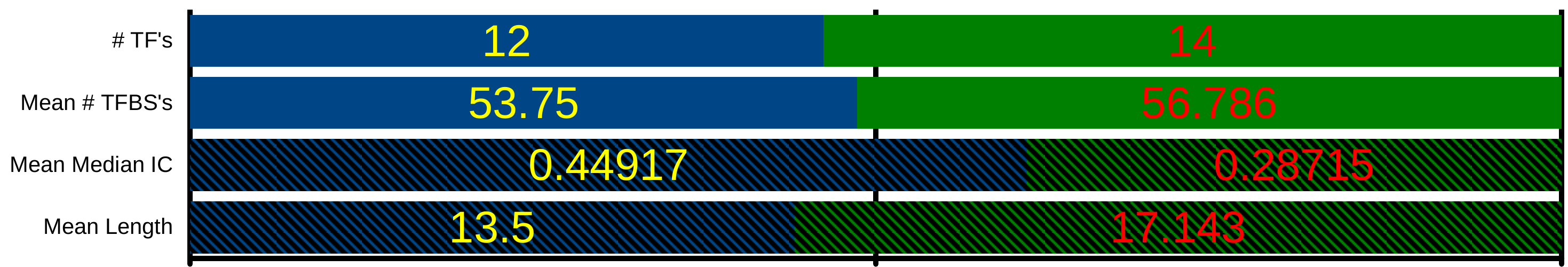}
\includegraphics[width = 0.5\textwidth]{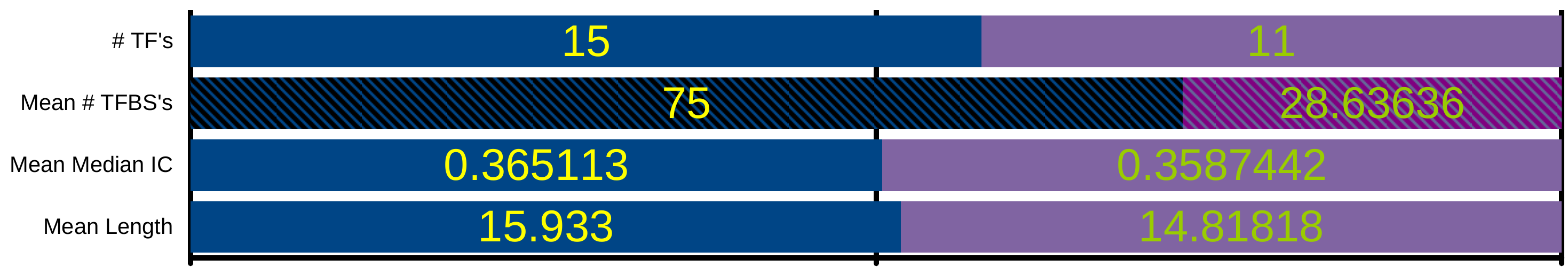}
\includegraphics[width = 0.5\textwidth]{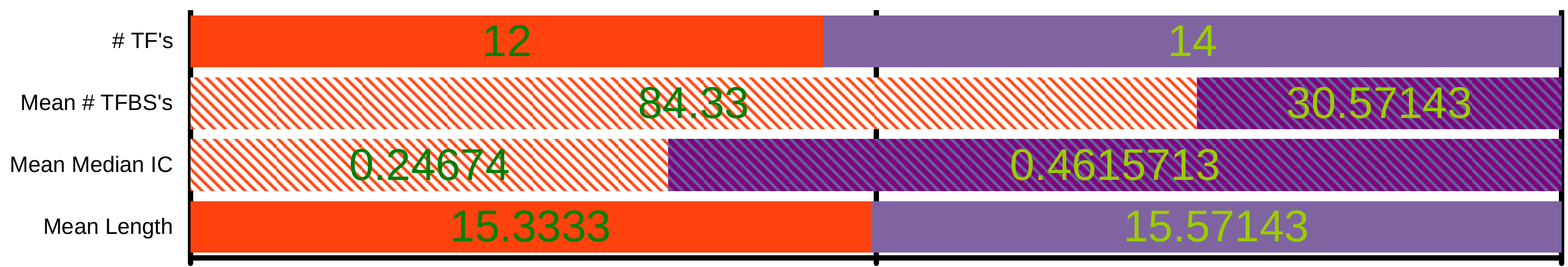}
\includegraphics[width = 0.5\textwidth]{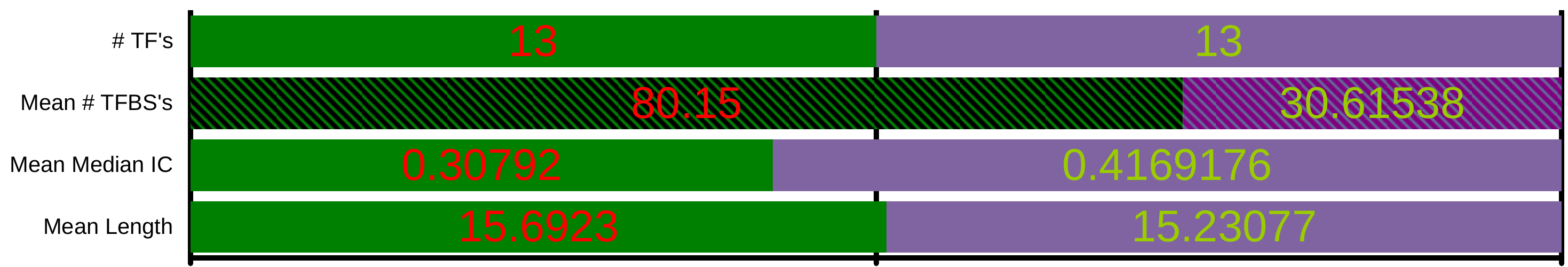}
\caption{Pair-wise comparisons of ODV\_P, 2-centroid\_P, PSSM\_P and ULPB without information content
on the second data set of 26 TF's.
Three factors except for \# TF's are tested for statistical significance. Significant factors are marked
by striped bars.}
\label{fig:pairwise_comp_Sal10}
\end{figure}

Comparisons were also made between the four comparable methods, ODV\_P, 2-centroid\_P, PSSM\_P and ULPB, on the
second data set of 26 TF's. Fig.~\ref{fig:pairwise_comp_Sal10} shows the bar plots. 
The plots suggest that 2-centroid\_P performs better than PSSM\_P when a TF has higher median IC and shorter binding sites.
2-centroid\_P performs better than ODV\_P when a TF has more known TFBS's, ODV\_P outperforms ULPB when a TF
has less known TFBS's and higher median IC, and ODV\_P performs better than PSSM\_P when a TF has less
known TFBS's.
%We also noticed that ULPB requires slightly more known TFBS's than the other methods.

\begin{figure}[!t]
\centering
\includegraphics[width = 0.5\textwidth]{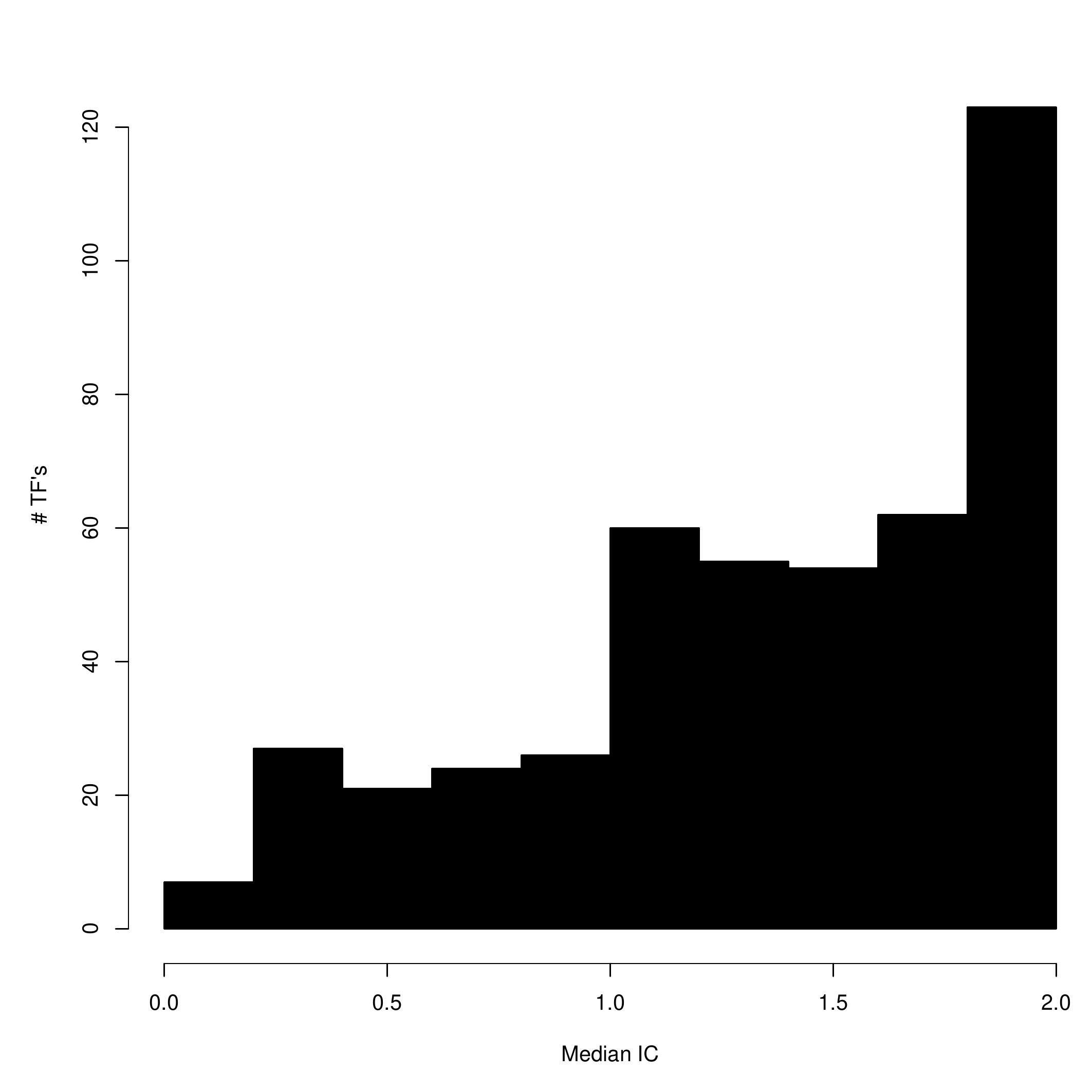}
\caption{Distribution of median IC of 459 eukaryotic transcription factors in the JASPAR database.}
\label{fig:hist_JASPAR}
\end{figure}

From the observations above, we can see that methods utilizing negative examples
%such as 2-centroid\_P and ODV\_P
tend to perform better on TF's with higher median information content.
% and shorter binding sites.
This suggests that the proposed 2-centroid and ODV methods are well-suited for identifying eukaryotic transcription factor
binding sites. Fig.~\ref{fig:hist_JASPAR} shows the distribution of median IC of 459 eukaryotic transcription factors
in the JASPAR database \cite{Bry08}, where 75\% (344 out of 459) of the TF's have median IC above 1.02.
According to our analysis shown in Fig.~\ref{fig:pairwise_comp}~and~\ref{fig:pairwise_comp_Sal10},
the 2-centroid and ODV methods perform significantly better than other compared methods when a TF
has relatively high median IC.

Moreover, properties revealed in Fig.~\ref{fig:pairwise_comp}~and~\ref{fig:pairwise_comp_Sal10} can potentially help
improve our 2-centroid and ODV methods. We can see in Fig.~\ref{fig:regression}
that the median information content of a TF can be as low as 0.05. We suspect that the motif of such TF
is actually a mixture of two or more motif subtypes, which contributes to its low median IC.
We expect the motif subtypes of a TF to have higher median IC. Thus, a method can first identify motif
subtypes contained in the known TFBS's of a TF and then search for individual subtypes.

\subsection{Motif Subtypes Improve the 2-centroid Method}

It has been shown that the binding sites of a TF can be better represented by 2 motif subtypes than by a single motif
\cite{Han05,Geo06}. In search for new binding sites, two position-specific scoring matrices are used to
score an $l$-mer and the higher score of the two is assigned to this $l$-mer.
Searching with two PSSM's was shown to be superior to searching with a single PSSM by cross-species conservation
statistics in these studies.

To validate our hypothesis proposed in Section~\ref{sec:pairwise_comp}, we coupled motif subtypes with the
centroid method as well as the 2-centroid method. Our approach to motif subtype identification is
slightly different from those in previous work \cite{Han05,Geo06}, while the idea is similar.
As usual, all the $l$-mers were first embedded in the Euclidean
space as described in Section~\ref{sec:centroid}. The known binding sites of a TF were clustered into two subtypes
by the $k$-means algorithm \cite{deH04}. The centroids of these two subtypes, $\bm{\mu}_{+1}$ and $\bm{\mu}_{+2}$,
were then computed. The centroid method coupled with motif subtypes is denoted by centroid\_C and it scores an $l$-mer $t$
by
\[
	\max\left\{\bm{\mu}_{+1}\transpose \bm t, \bm{\mu}_{+2}\transpose \bm t \right\},
\]
where $\bm t$ denote the $l$-mer $t$ embedded in the Euclidean space.
On the other hand, the 2-centroid method coupled with motif subtypes is denoted by 2-centroid\_C and it score
an $l$-mer $t$ by
\[
	\max\left\{\left(\bm{\mu}_{+1} - \bm{\mu}_- \right)\transpose \bm t, \left(\bm{\mu}_{+2} - \bm{\mu}_- \right)\transpose \bm t \right\},
\]
where $\bm{\mu}_-$ is the centroid of the non-binding sites.

\begin{figure*}[!t]
\centering
\includegraphics[width=\textwidth]{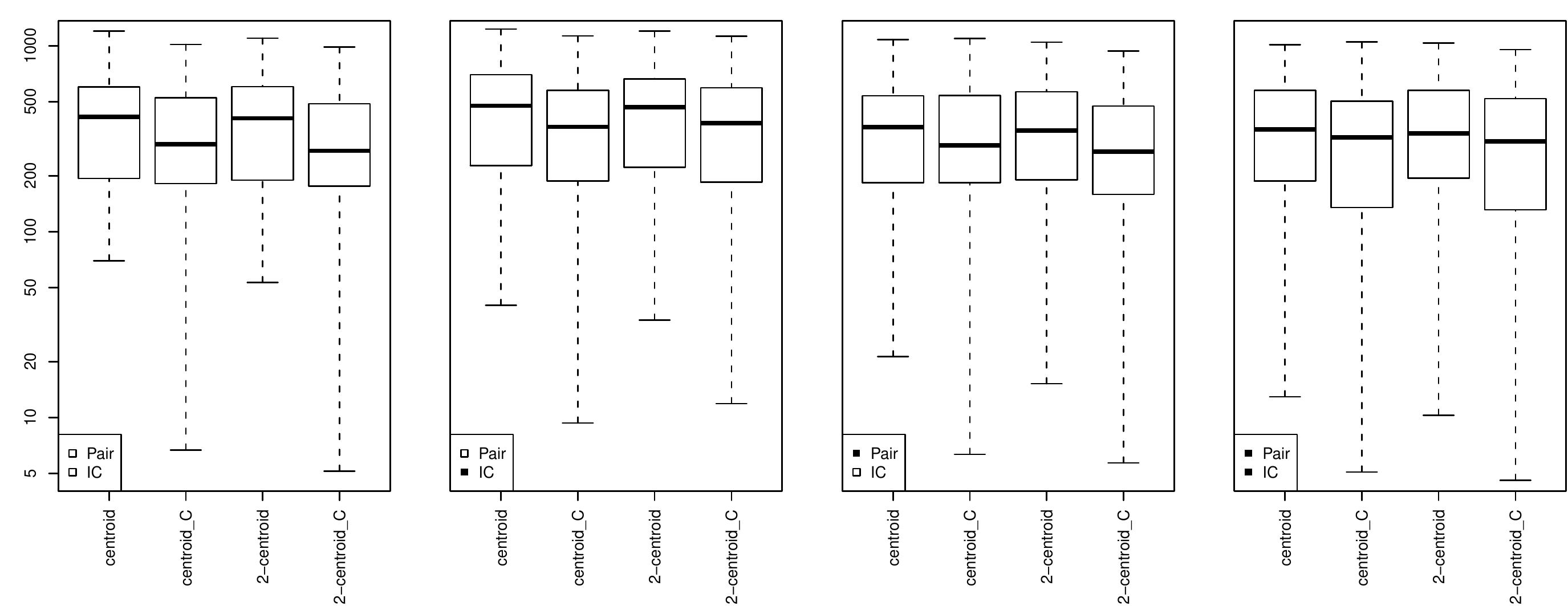}
\caption{Box plots showing the LOO CV results of methods centroid, centroid\_C, 2-centroid and 2-centroid\_C.
Pair denotes the use of nucleotide pairs and IC indicates weighting nucleotides and nucleotide pairs with
information content.}
\label{fig:boxplot_subtypes}
\end{figure*}

We assessed and compared centroid\_C and 2-centroid\_C to their counterparts without motif subtypes by leave-one-out cross-validation
on the second data set of 26 TF's. Results summarized as box plots are shown in Fig.~\ref{fig:boxplot_subtypes}, where
Pair denotes the use of nucleotide pairs and IC indicates weighting nucleotides and nucleotide pairs with information content.
In all the four cases, significant improvement was observed when motif subtypes were taken into account.
Table~\ref{tab:subtype} elucidates the impact of motif subtype identification on our 2-centroid method.
The first column shows that, before introducing motif subtypes, the improvement of 2-centroid over centroid is only
statistically significant in the first row. The second column displays significant improvement of centroid\_C over centroid, which
was anticipated and consistent with the results reported in \cite{Han05,Geo06}.
The third column shows significant improvement of 2-centroid\_C over 2-centroid in all four cases.
We observed that the improvement of 2-centroid\_C over 2-centroid is always more significant than
the improvement of centroid\_C over centroid. This implies that our 2-centroid method benefitted even more from
the identification of motif subtypes. The last column indicates that, after the introduction of motif subtypes, 
2-centroid\_C significantly outperforms centroid\_C in all cases. These results confirmed our hypothesis that,
for TF's with low median IC, methods employing non-binding sites should be coupled with motif subtype identification.

\begin{table*}[!t]
	\centering
	\begin{threeparttable}
	%% increase table row spacing, adjust to taste
	\renewcommand{\arraystretch}{1.3}
	% if using array.sty, it might be a good idea to tweak the value of
	% \extrarowheight as needed to properly center the text within the cells
	\caption{Improvement by Identifying Motif Subtypes}
	\label{tab:subtype}
	%% Some packages, such as MDW tools, offer better commands for making tables
	%% than the plain LaTeX2e tabular which is used here.
	\begin{tabular}{cc|rr|rr|rr|rr}
	& & \multicolumn{2}{c|}{2-centroid $\rightarrow$ centroid} & \multicolumn{2}{c|}{centroid\_C\tnote{e} $\rightarrow$ centroid} &
	\multicolumn{2}{c}{2-centroid\_C $\rightarrow$ 2-centroid} & \multicolumn{2}{c}{2-centroid\_C $\rightarrow$ centroid\_C}\\
	Pair\tnote{a} & IC\tnote{b} & \# better\tnote{c} & $p$-value\tnote{d} & \# better & $p$-value & \# better & $p$-value & \# better & $p$-value\\
	\hline\hline
	$\Box$ & $\Box$ & 19 & $2.793 \times 10^{-2}$ &  18 & $5.093 \times 10^{-3}$ & 21 & $2.205 \times 10^{-5}$ & 21 & $1.205 \times 10^{-3}$\\
	$\Box$ & $\blacksquare$ & 18 & $\bm{ 5.037 \times 10^{-2} }$ & 19 & $3.727 \times 10^{-4}$ & 22 & $1.135 \times 10^{-5}$ & 19 & $5.983 \times 10^{-3}$\\
	$\blacksquare$ & $\Box$ & 17 & $\bm{ 9.937 \times 10^{-2} }$ & 16 & $3.757 \times 10^{-2}$ & 23 & $6.661 \times 10^{-6}$ & 18 & $2.806 \times 10^{-3}$\\
	$\blacksquare$ & $\blacksquare$ & 17 & $\bm{ 1.185 \times 10^{-1} }$ & 17 & $7.003 \times 10^{-3}$ & 20 & $2.325 \times 10^{-4}$ & 19 & $8.807 \times 10^{-3}$\\\hline
	\end{tabular}
	\begin{tablenotes}
		\item [a] Whether a method uses nucleotide pairs.
		\item [b] Whether a method weights nucleotide and nucleotide pairs with information content.
		\item [c] The number of TF's supporting the relationship being tested.
		\item [d] $p$-value of the relationship produced by a statistical test \cite{Wil45}.
		\item [e] Suffix \_C denotes coupling a method with motif subtypes.
	\end{tablenotes}
\end{threeparttable}
\end{table*}

\begin{figure*}[!t]
\centering
\subfloat[{ }]{\label{fig:FlhDC_logos_before}
\includegraphics[width=0.33\textwidth]{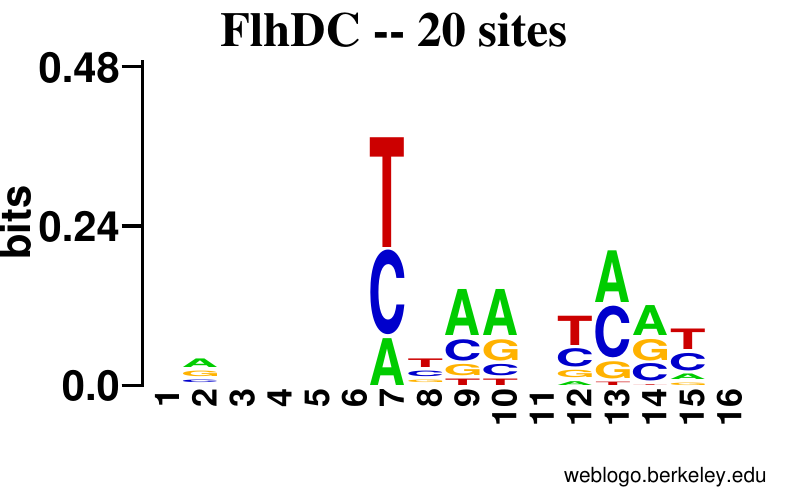}
}              
\subfloat[{ }]{\label{fig:FlhDC_logos_after}
\includegraphics[width=0.33\textwidth]{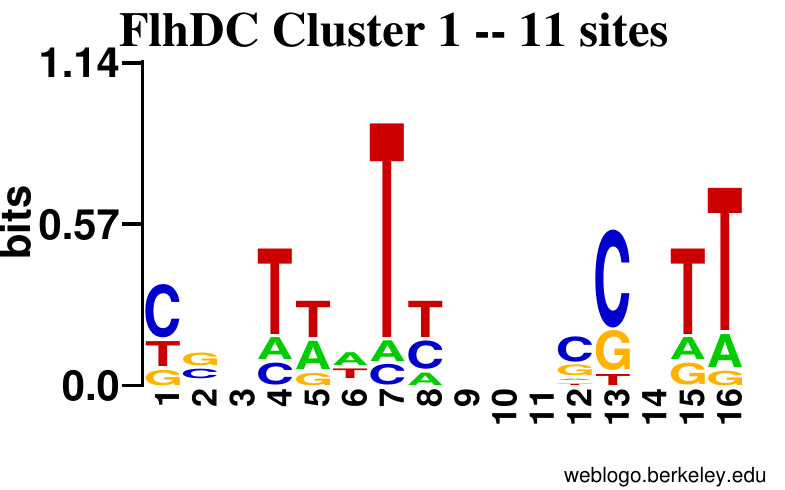}
\includegraphics[width=0.33\textwidth]{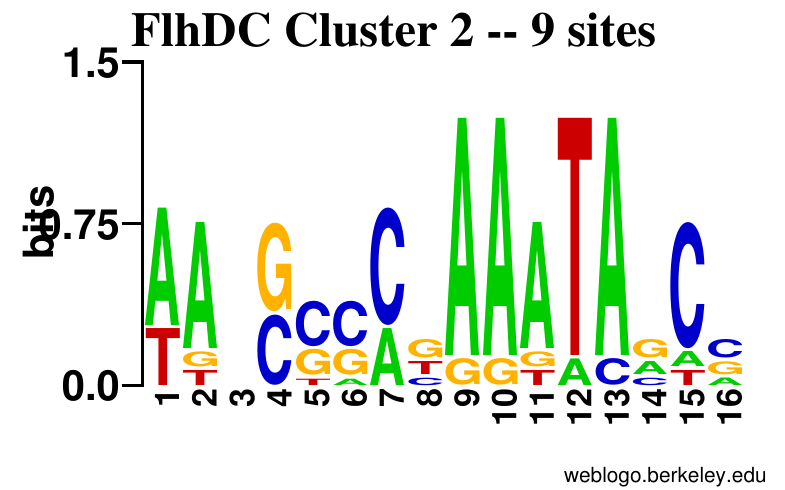}
}\\
\subfloat[{ }]{\label{fig:FlhDC_scatter}\includegraphics[width=0.5\textwidth]{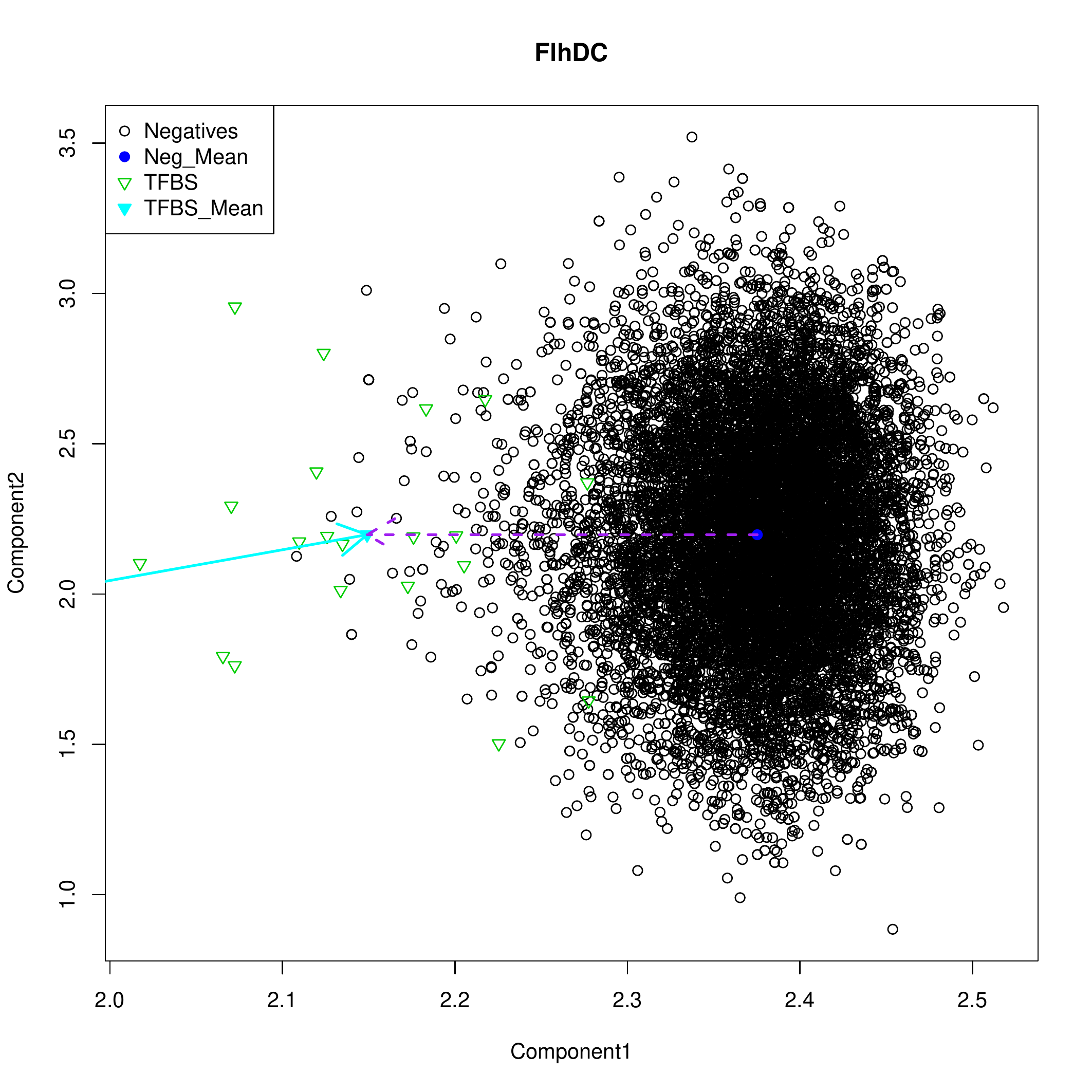}}
\subfloat[{ }]{\label{fig:FlhDC_clust_scatter}\includegraphics[width=0.5\textwidth]{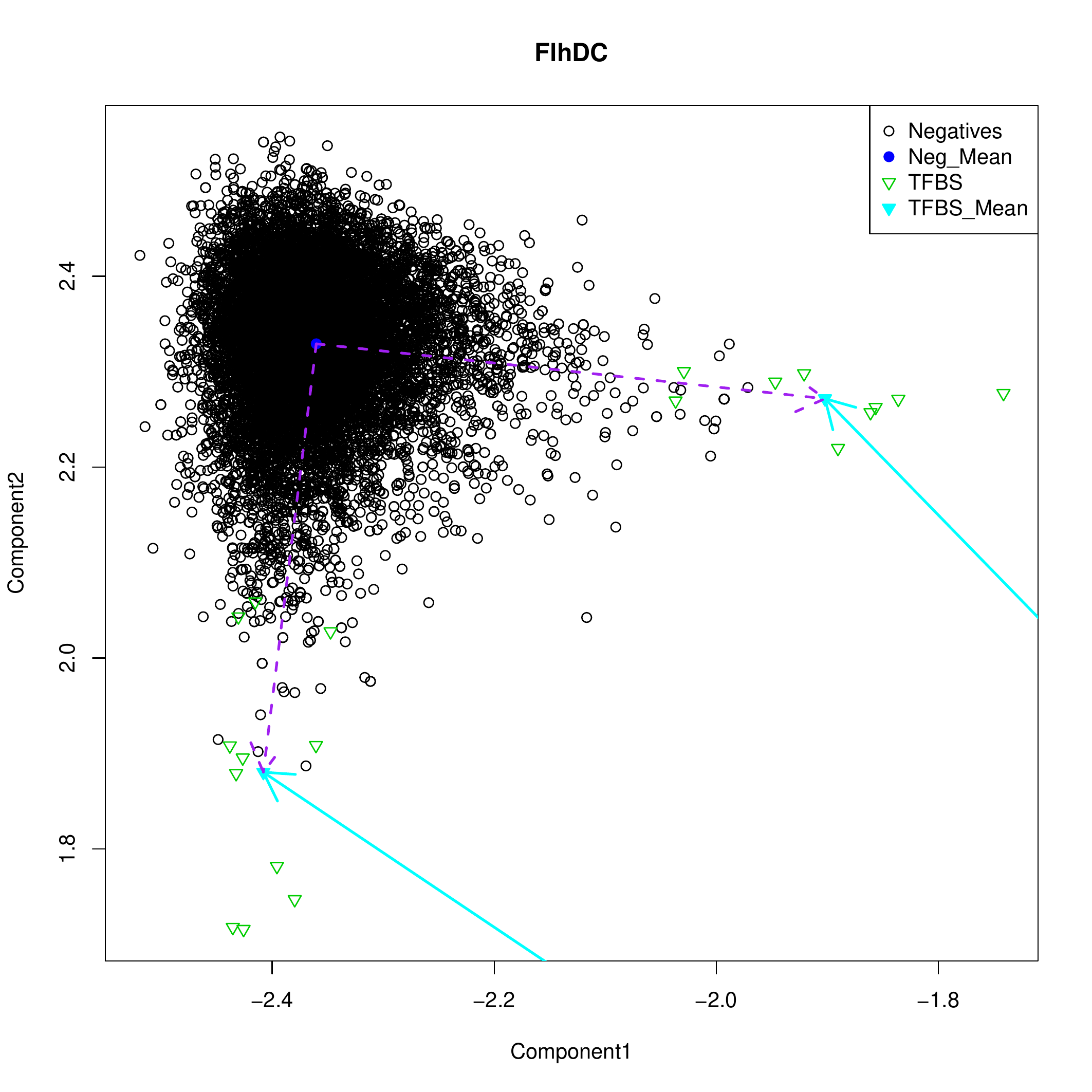}}
\caption{
%Sequence logos of FlhDC and scatter plots of binding and non-binding sites of FlhDC.
Illustration of the 2-centroid\_C method with nucleotide pairs on transcription factor FlhDC in the second data set.
Axes in (c) and (d) were found by Fisher's discriminant analysis \cite{Has09}.
(a) Sequence logo before motif subtype
identification. (b) Sequence logos of two motif subtypes identified by $k$-means clustering. (c) Scatter plot of binding sites,
non-binding sites and their respective centroids, $\bm \mu_+$ and $\bm \mu_-$. The solid arrow identifies the vector $\bm \mu_+$,
while the dashed arrow denotes the vector $\bm \mu_+ - \bm \mu_-$. (d) Scatter plot of two clusters of binding sites,
non-binding sites and their respective centroids, $\bm \mu_{+1}, \bm \mu_{+2}$ and $\bm \mu_-$.
The two solid arrows represent vectors $\bm \mu_{+1}$ and $\bm \mu_{+2}$, while the two dashed arrows denotes vectors
$\bm \mu_{+1} - \bm \mu_-$ and $\bm \mu_{+2} - \bm \mu_-$.}
\label{fig:FlhDC_P}
\end{figure*}

Fig.~\ref{fig:FlhDC_P} illustrates the application of 2-centroid\_C with nucleotide pairs
to transcription factor FlhDC in the second data set. It can be seen in Fig.~\ref{fig:FlhDC_logos_before}
that the information content of FlhDC is low at all the 16 positions. After motif subtype identification,
the two subtypes display distinct patterns and the information content of the two subtypes was greatly
improved as seen in Fig.~\ref{fig:FlhDC_logos_after}. Fig.~\ref{fig:FlhDC_scatter}
shows a scatter plot of binding sites, non-binding sites and their respective centroids, while Fig.~\ref{fig:FlhDC_clust_scatter}
shows a scatter plot of binding sites belonging to two subtypes, non-binding sites and their respective centroids
after motif subtype identification. Many binding sites are not distinguishable from non-binding sites in Fig.~\ref{fig:FlhDC_scatter}.
However, after motif subtype identification, TFBS's became separable from non-TFBS's as seen in Fig.~\ref{fig:FlhDC_clust_scatter},
resulting in 1.7-fold improvement in average rank.

\subsection{Connection between ODV and PSSM/ULPB}\label{sec:connection}

Finally, we elucidate the relation between ODV and PSSM/ULPB.
We first derive the connection between the optimal discriminating vector method and the position-specific scoring
matrix method. Without loss of generality, we do not include nucleotide pairs in the derivation for simplicity reasons.
We abuse notations for a moment and let $\beta_i(\text A) = \beta_{4i-3}$, $\beta_i(\text C) = \beta_{4i-2}$,
$\beta_i(\text G) = \beta_{4i-1}$ and $\beta_i(\text T) = \beta_{4i}$. (\ref{eq:scoring_function})
then becomes
\begin{align}
	\bm \beta\transpose \bm t &= \sum_{i=1}^l \beta_i(t_i)\sqrt{w_i}
	= \sum_{i=1}^l \log\left( \frac{f_i(t_i)k_i}{f(t_i)} \right) w_i\notag\\
	&= \sum_{i=1}^l \log\left( \frac{f_i(t_i)}{f(t_i)} \right) w_i + \sum_{i=1}^lw_i\log k_i,\label{eq:ODV_PSSM}
\end{align}
where $f_i(t_i) = \frac{1}{k_i}\exp\left(\frac{\beta_i(t_i)}{\sqrt{w_i}}\right)f(t_i)$ is the
position-specific nucleotide frequency for $t_i$ induced by $\beta_i(\cdot)$ and 
\[
k_i = \sum_{u \in \{\text{A, C, G, T}\}} \exp\left(\frac{\beta_i(u)}{\sqrt{w_i}}\right)f(u) > 0
\]
is a scaling factor for position $i$ since ODV does not impose the constraints
$\sum_{u \in \{\text{A, C, G, T}\}} f_i(u) = 1, \,\,\forall i$. 
From (\ref{eq:ODV_PSSM}),
we note that $\sum_{i=1}^lw_i\log k_i$ does not depend on $t$ and thus $\bm \beta$ is optimal if and only
if $\{f_i(u)|u \in \{\text{A, C, G, T}\} \text{ and }i = 1, 2, \ldots, l\}$, is optimal. Therefore, an optimal PSSM
can be obtained from our ODV method.

The ungapped likelihood under positional background method is similar to the PSSM\_P method in that
both methods score nucleotides and nucleotide pairs. The ULPB method scores a $l$-mer $s$ by looking at
the first nucleotide $s_1$ and all the $l-1$ adjacent nucleotide pairs $s_1s_2, s_2s_3, \ldots, s_{l-1}s_l$.
Therefore, we can embed $s$ in $\mathbb R^{20l - 16}$ by transforming $s_1$ into 4 dummy variables and
each of the $l-1$ pairs into 16 dummy variables as described in Section~\ref{sec:centroid}.
An optimal discriminating vector $\bm \beta \in \mathbb R^{20l - 16}$ can then be found by applying
our ODV method described in Section~\ref{sec:ODV}. Following similar arguments, we can see that
there is a one-to-one correspondence between elements of $\bm \beta$ and
$\left\{f_1(u), f_i(v|u)| u, v \in \{\text{A, C, G, T}\} \text{ and } i = 1, 2, \ldots, l-1 \right\}$
in (\ref{eq:ULPB}). Hence, an optimal ULPB can also be obtained from our ODV method.

\begin{figure}[!t]
\centering
\includegraphics[width = 0.4\textwidth]{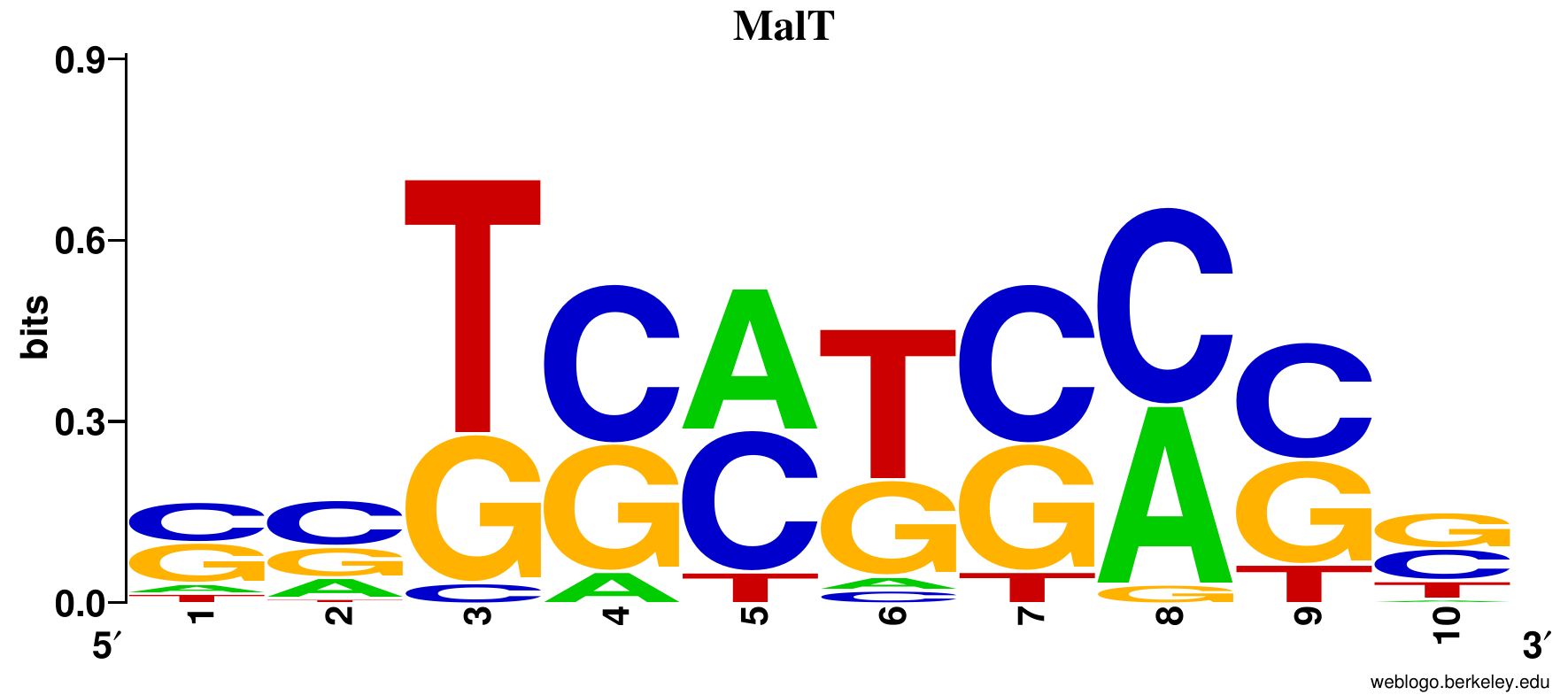}
\includegraphics[width = 0.4\textwidth]{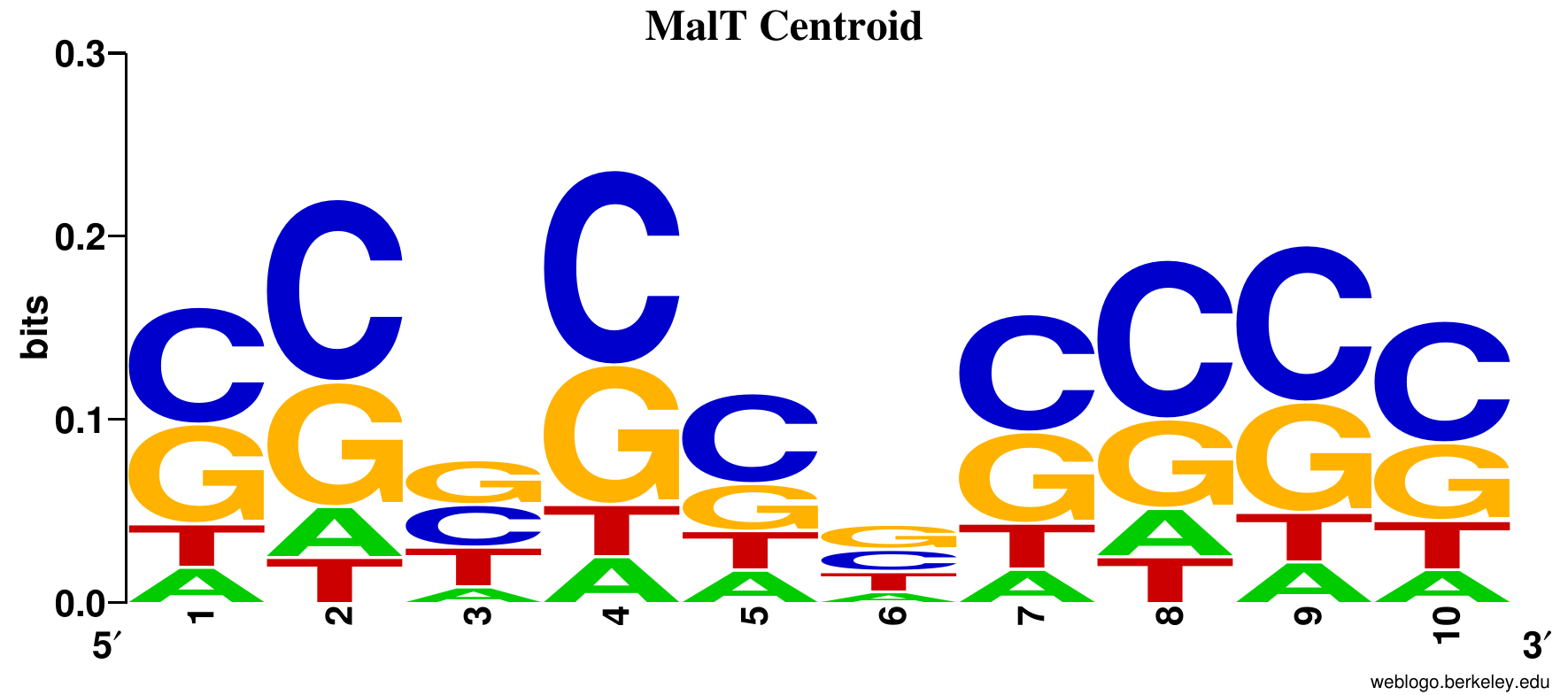}
\caption{Two sequence logos of TF MalT. Top: PSSM; Bottom: centroid.}
\label{fig:logo}
\end{figure}

One direct implication of the connection established above is that
a vector obtained by the centroid, 2-centroid or ODV methods can be compared to a PSSM model
in the same framework. As an example, Fig.~\ref{fig:logo} shows two sequence logos \cite{Cro04}
of TF MalT in the second data set. The top logo represents the signature of the known binding
sites, while the bottom one is obtained by converting the centroid $\bm \mu_+$ to a PSSM model
as in (\ref{eq:ODV_PSSM}) with $\bm \beta = \bm \mu_+$. The two logos display distinct
patterns of the two methods, implying difference in performance. The PSSM method gave an
average rank of 233.9, while the centroid method gave an average rank of 69.8.
Clearly, the performance difference lies in the difference between the two logos.
We can see that the two logos are very different at positions 3, 5, 6 and 10.
Position 3 indicates that down-weighting letter T results in better performance.
Position 10 shows that the influence of letter A is underestimated in the PSSM model.
Other positions can be similarly compared and interpreted as well.

\section{Conclusion}\label{sec:conclusion}
% REVISE:
% Impact on Eukaryotic TFBS's

In this work, we investigated the use of negative examples in the TFBS search problem. To utilize
negative examples, we proposed the 2-centroid and ODV methods, which are natural extensions of
the centroid method. The proposed methods were compared to state-of-the-art methods relying purely
on positive examples as well as a method considering negative examples.
Comprehensive LOO CV results showed that non-TFBS's are indeed helpful for TFBS search. The large
number of non-binding sites can be significantly reduced by sampling a small representative set
by LOO CV.

Not surprisingly, there is no single best TFBS search method or similarity measure for all the TF's.
The best combination of similarity measure and search method can be found for a particular TF by
CV experiments. Nevertheless, pair-wise comparisons between methods revealed interesting properties of
methods compared in this work. In particular, we showed that the 2-centroid and ODV methods are
significantly better than the other methods when a TF has relatively high median information content.
Even for TF's with low median information content, preceded by motif subtype identification, the
2-centroid method was shown to be effective in searching for binding sites belonging to individual
subtypes. The ODV method can be easily coupled with motif subtype identification as well and we
believe significant improvement can be expected.

All the experiments in this work were conducted on prokaryotic transcription factors, i.e.,
TF's in the \textit{E. coli} K-12 genome. We claim that the proposed 2-centroid and ODV are well-suited for
eukaryotic transcription factor binding site search as well. This is based on characteristics of the
proposed methods and summary statistics of 459 eukaryotic transcription factors in the JASPAR database.
Finally, we derived the connection between our ODV method and the PSSM method,
showing that an optimal vector in ODV implies an optimal scoring matrix in PSSM and vice versa.
Properly embedding an $l$-mer in an Euclidean space, the same connection between ODV and ULPB
can be established as well.

% FUTURE WORK: HANDLE KNOWN BINDING SITES OF VARIABLE LENGTHS
The effects of negative examples on eukaryotic transcription factor binding site search will be investigated.
Our future work also aims for extending our proposed methods to handling known binding sites of variable
lengths. We will seek to approach this problem without resorting to multiple sequence alignment,
which is notoriously time-consuming.
In the meantime, we will also seek to identify better similarity measures than those investigated in this study.

% if have a single appendix:
%\appendix[Proof of the Zonklar Equations]
% or
%\appendix  % for no appendix heading
% do not use \section anymore after \appendix, only \section*
% is possibly needed

% use appendices with more than one appendix
% then use \section to start each appendix
% you must declare a \section before using any
% \subsection or using \label (\appendices by itself
% starts a section numbered zero.)
%

%\appendices
%\section{Proof of the First Zonklar Equation}
%Appendix one text goes here.

% you can choose not to have a title for an appendix
% if you want by leaving the argument blank
%\section{}
%Appendix two text goes here.

% use section* for acknowledgement
\ifCLASSOPTIONcompsoc
  % The Computer Society usually uses the plural form
  \section*{Acknowledgments}
\else
  % regular IEEE prefers the singular form
  \section*{Acknowledgment}
\fi

This work was supported in part by National Science Foundation [grant number CCF-0755373].

% Can use something like this to put references on a page
% by themselves when using endfloat and the captionsoff option.
\ifCLASSOPTIONcaptionsoff
  \newpage
\fi

% trigger a \newpage just before the given reference
% number - used to balance the columns on the last page
% adjust value as needed - may need to be readjusted if
% the document is modified later
%\IEEEtriggeratref{8}
% The "triggered" command can be changed if desired:
%\IEEEtriggercmd{\enlargethispage{-5in}}

% references section

% can use a bibliography generated by BibTeX as a .bbl file
% BibTeX documentation can be easily obtained at:
% http://www.ctan.org/tex-archive/biblio/bibtex/contrib/doc/
% The IEEEtran BibTeX style support page is at:
% http://www.michaelshell.org/tex/ieeetran/bibtex/
\bibliographystyle{IEEEtran}
% argument is your BibTeX string definitions and bibliography database(s)
\bibliography{IEEEabrv,tfbs}

% biography section
% 
% If you have an EPS/PDF photo (graphicx package needed) extra braces are
% needed around the contents of the optional argument to biography to prevent
% the LaTeX parser from getting confused when it sees the complicated
% \includegraphics command within an optional argument. (You could create
% your own custom macro containing the \includegraphics command to make things
% simpler here.)
%\begin{biography}[{\includegraphics[width=1in,height=1.25in,clip,keepaspectratio]{mshell}}]{Michael Shell}
% or if you just want to reserve a space for a photo:

\begin{IEEEbiography}{Chih Lee}
\end{IEEEbiography}

\begin{IEEEbiography}{Chun-Hsi Huang}
\end{IEEEbiography}

% if you will not have a photo at all:
%\begin{IEEEbiographynophoto}{John Doe}
%Biography text here.
%\end{IEEEbiographynophoto}

% insert where needed to balance the two columns on the last page with
% biographies
%\newpage

%\begin{IEEEbiographynophoto}{Jane Doe}
%Biography text here.
%\end{IEEEbiographynophoto}

% You can push biographies down or up by placing
% a \vfill before or after them. The appropriate
% use of \vfill depends on what kind of text is
% on the last page and whether or not the columns
% are being equalized.

%\vfill

% Can be used to pull up biographies so that the bottom of the last one
% is flush with the other column.
%\enlargethispage{-5in}

% that's all folks
\end{document}